\documentclass[11pt,aps,preprint,superscriptaddress,nofootinbib]{revtex4-1}
\usepackage[active]{srcltx}
\usepackage[utf8]{inputenc}
\usepackage{latexsym}
\usepackage{hyperref}
\usepackage{amsmath}
\usepackage{amsfonts}
\usepackage{graphicx}
\usepackage{slashed}
\usepackage{xcolor}
\usepackage{soul}
\usepackage{slashed}
\usepackage{braket}
\usepackage{natbib}

\begin{document}
\title{Lattice Clifford fractons and their Chern-Simons-like theory}
\author{Weslei B. Fontana}
\email{weslei@uel.br}
\affiliation{Physics Department, Boston University, Boston, MA, 02215, USA}
\affiliation{Departamento de F\'isica, Universidade Estadual de Londrina, 86057-970, Londrina, PR, Brasil}

\author{Pedro R. S. Gomes}
\email{pedrogomes@uel.br}
\affiliation{Departamento de F\'isica, Universidade Estadual de Londrina, 86057-970, Londrina, PR, Brasil}

\author{Claudio Chamon}
\email{chamon@bu.edu}
\affiliation{Physics Department, Boston University, Boston, MA, 02215, USA}
	
	
\begin{abstract}
	
{\bf We use Dirac matrix representations of the Clifford algebra to build
	fracton models on the lattice and their effective Chern-Simons-like
	theory. As an example, we build lattice fractons in odd $D$ spatial
	dimensions and their $(D+1)$ spacetime dimensional effective theory. The model possesses an
	anti-symmetric $K$ matrix resembling that of hierarchical quantum Hall
	states. The gauge charges are conserved in sub-dimensional manifolds
	which ensures the fractonic behavior. The construction extends to any
	lattice fracton model built from commuting projectors and with tensor
	products of spin-$1/2$ degrees of freedom at the sites.
}
\end{abstract}
	
	
\maketitle

\section{Introduction}
A major goal of condensed matter physics is to understand and to
classify all possible phases of matter; another one is to uncover
phases outside contemporary paradigms. While these two goals are
evidently contradictory, together they move the field forward. An
example of a new class of systems whose complete understanding is
still in progress is that of what is now commonly referred to as
fractons in general, or more precisely, systems with fracton
excitations. These systems have peculiar properties, including ground
state degeneracies that depend both on topology and geometry of
lattice discretizations, and excitations with restricted mobility
that, in turn, make the dynamical relaxation to the ground states
slow~\cite{Chamon2005,Terhal2011,Haah2011,Castelnovo2011,Haah2015}.

Recent reviews of fractons can be found in
\cite{NandkishoreHermele2019} and \cite{PretkoChenYou2020}. Thus far,
they are classified into two types: in Type I phases a single fracton
excitation cannot move alone, but a pair can bind into mobile dipoles;
in Type II phases, all excitations are immobile
\cite{Haah2011,Haah2015,Yoshida2013,Vijay2016}. It is this inherent
immobility of isolated excitations that lead to slow dynamical
behavior \cite{Chamon2005,Nandkishore2017,Pai2019}. The same
restricted mobility and slow dynamical relaxation of excitations might
be useful for building quantum memories
\cite{Bravyi2013,Terhal2015,Brown2019}. In addition, fractons possess
connections to elasticity theory \cite{Radzihovsky2018, Gromov2019a}
and gravity \cite{Pretko2017}.

Fracton phases were originally constructed in lattice models; while
their peculiar properties might appear unnatural for continuum
descriptions, the construction of effective field theories that
capture their low-energy properties is possible, as shown by Slagle
and Kim~\cite{SlagleKim2017} in the X-cube
model~\cite{castelnovo2010,Haah2015}. The construction of effective
field theories enables much further progress~\cite{SlagleKim2017,
	Burnell2019,Radicevic2020, Chen2019,
	Williamson2019,Gromov2019,Bulmash2018}. Some features of fracton
excitations are captured by the field theories in simple
ways. Restricted mobility, for example, is encoded in additional
charge conservation laws along sub-dimensional manifolds, such as
planes for 3-dimensional (3D)\footnote{When referring to the dimensionality of the spacetime in this work we will use the notation $(D+1)$, with $D$ the number of spatial components and the $+1$ refers to the time direction.} models, besides the conservation of
total charge in the whole volume. The conservation of charges in
planes implies that a dipole in the perpendicular direction is
conserved. Hence charge conservation in sub-manifolds is equivalent to
the conservation of vector charges (dipoles), a feature of higher-rank
gauge theories \cite{Pretko2017a, Pretko2017b,
	Xu2006,Rasmussen2016,Wen2012,Wen2006,Horava2010, Vijay2020,
	Wang2019, Seiberg2020, Burnell2018, Seiberg2020a, Seiberg2020b,
	Seiberg2020c}, which, in general, are gapless. Nevertheless, gapped
fracton models can be obtained from higher-rank gauge theories via the
Higgs mechanism \cite{Chen2018,Bulmash2018a}. Gapped 3D fractons can
also be obtained by either stacking \cite{Shirley2018, Shirley2019,
	Wang2019, Schmitz2019, Chen2019a, Fuji2019, Wang2019a} or glueing
\cite{Wen2020, Aasen2020, Wang2020} known $(2+1)$-dimensional topological
orders.

You {\it et al.} present a different route to a fracton field theory
that is not cast as a higher-rank gauge model. They present a
Chern-Simons-like action with vector gauge fields that contains the
sub-manifold conservation laws, hence also conserving dipoles. Their
theory is gapped, and it can be discretized to a lattice to arrive at
the Chamon model of Ref.~\cite{Chamon2005}. The connection to Chern-Simons-like
theories is appealing in that one would hope they can be generalized
to describe classes of gapped fractons, much like Chern-Simons
theories can describe classes of quantum Hall states~\cite{Wen1995}.

In this work, we construct families of Chern-Simons-like theories of
gapped fractons. These theories have multiple gauge charges, and are
described by an anti-symmetric $K$ matrix and associated charge
vectors. We arrive at these theories starting from microscopic lattice
models, where we place a number $n$ of spin-1/2 degrees of freedom (or
qubits) at the sites. Such starting point is rather generic, and
encompasses models such as the Chamon and Haah codes. Instead of
tensor products of Pauli operators, we use the Dirac representation of
the Clifford algebra to describe the site degrees of freedom. We show
that the Dirac representation with $2^n$-dimensional matrices is a
natural mathematical framework to build the lattice models, and makes
the connection to the field theory, a bosonization of sorts, rather
simple. In the lattice theory, the fracton nature of the models are
simple consequences of the lattice connectivity and the Clifford
algebra, for example the immobility of single defects. In the
continuum theory, these properties translate into charge conservation
laws in sub-manifolds.

For the sake of giving a concrete but yet general example of the
construction of these Clifford fractons, we build fracton models in
any odd $D=2n+1$ spatial dimensions. This example allows one to track more
easily the use of the $2^n\times 2^n$ anti-commuting Dirac matrices
$\gamma^I$ with $I=1, 2,\dots, 2n+1$, where
$\gamma^1\,\gamma^2\,\ldots\,\gamma^{2n+1}=i^{n}$. The model of
Ref.~\cite{Chamon2005} corresponds to the simplest case, with $D=3$ and
$2\times 2$ representations of the Dirac matrices. The $2n+1$ Dirac
matrices form a maximal set of anti-commuting operators, and no
operator (any product of Dirac matrices) other than the identity
commutes with less than two of the $\gamma$'s; it is this algebraic
property that impedes the propagation of single fracton excitations.

We encode the anti-commutation relations of the Dirac matrices in a
$2n\times 2n$ anti-symmetric matrix $K$ for constructing a model in
the continuum. This ``bosonization''-type scheme is a generalization
of that in Ref.~\cite{SlagleKim2017}. The generic bosonic formulation
in terms of $K$ matrices and charge vectors allows us to take a
continuum limit, and arrive at a (D+1)-dimensional Chern-Simons-like action
\begin{align}
{\cal L} &=\sum_{a,\,b=1}^{2n}
\frac{1}{2\pi}\;K_{ab}\;A_a \,\partial_0 \, A_b
+
\frac{1}{\pi}\;\sum_{\alpha} \;K_{ab}\;A^{(\alpha)}_0\; \mathcal{D}^{(\alpha)}_a\,A_b
,
\end{align}
where the differential
$\mathcal{D}^{(\alpha)}_a=\sum_{I=1}^{D}T^{(I,\,\alpha)}_a\,\partial^2_{I}$ operators are tied to charge vectors
$T^{(I,\,\alpha)}$ dictated by products of Dirac matrices in the
microscopic lattice theory. The lattice model also determines the
number of conserved currents that are minimally coupled to the $n$
fields $A^{(\alpha)}_0$, indexed by $\alpha=1,\dots,n$.  The action is
invariant under the $n$ gauge transformations
\begin{align}
& A_a
\rightarrow
A_a + \sum_{\alpha}\mathcal{D}^{(\alpha)}_a \zeta^{(\alpha)}
\;,
\\
& A^{(\alpha)}_0
\rightarrow
A^{(\alpha)}_0 +\partial_0 \zeta^{(\alpha)}
\nonumber
\;,
\end{align}
if $K_{ab}\;\mathcal{D}^{(\alpha)}_a\;\mathcal{D}^{(\beta)}_b=0$,
again a condition ensured by relations between the microscopic lattice
charge vectors $T^{(I,\,\alpha)}$ and the $K$ matrix. The conservation
of the $n$ currents
\begin{align}
\partial^{\; }_0\, J^{(\alpha)}_0 = \mathcal{D}^{(\alpha)}_a J^{\;}_a
\end{align}
not only on the full volume, but also on sub-manifolds, follows from a
linear dependence of the charge vectors, $\sum_{I=1}^D\;
T^{(I,\,\alpha)} = 0$, which in turn affects the differential operators
$\mathcal{D}^{(\alpha)}_a$.

These fracton models have a ground state degeneracy
\begin{align}
\textrm{GSD}
=
\left[2^{\frac{D-1}{2}}\;\text{Pf}\;(K)\right]^{2^{D-2}\;L}
\end{align}
for systems of linear size $L$ (hypervolume $L^D$). In the case of the
``integer'' fractons we constructed on the lattice, the $K$-matrix
Pfaffian equals 1, and the degeneracy is $2^{(D-1)\;2^{D-3}\;L}$.

The paper is organized as follows. In Sec. \ref{microscopic model}, we
construct, as example, a microscopic Clifford fracton model in
$D=2n+1$ spatial dimensions. In Sec. \ref{effective theory}, we
construct the corresponding effective theory in the continuum. In
Sec. \ref{Properties} we examine several properties of the effective
field theories. We close in Sec.\ref{conclusions}, with a brief
summary and final remarks. Details of several computations as well as
additional relevant discussions are presented in the appendices.


\section{Microscopic Model  in Arbitrary Odd Dimensions \label{microscopic model}}
The 3D model of Ref.~\cite{Chamon2005} uses the simplest representation of the
Clifford algebra, where the Dirac $\gamma$ matrices are $2\times 2$:
$\gamma^I=\sigma^I$, $I=1,2,3$, with the $\sigma^I$ the Pauli
matrices. The corresponding Hilbert space of the local degrees of
freedom is 2-dimensional.

In 5D, for example, we can use the $4\times 4$ representation of the
Clifford algebra, with the 5 Dirac matrices $\gamma^I,
I=1,2,3,4,5$. (We work in Euclidean space, so we list the matrices
from 1 to 4 plus the $\gamma^5$.) These matrices all anti-commute,
$\{\gamma^I,\gamma^J\}=2\delta_{IJ}$, and
$\gamma^1\gamma^2\gamma^3\gamma^4\gamma^5=i^2$. The local Hilbert
space is 4-dimensional in this case. (This representation is obtained
from a tensor product of two sets of Pauli matrices.)  In $D=2n+1$
dimensions, we work with $2^n\times 2^n$ representations of the
Clifford algebra, i.e. the Dirac matrices $\gamma^I, I=1,\dots, D$,
satisfying $\prod_{I=1}^D \gamma^I= i^n$. (We build these matrices
explicitly in appendix \ref{AA}.)

The construction of the fractons in odd-dimensional $D=2n+1$ space
proceeds as follows. We start with an face-centered hypercubic
lattice, that can be thought as the even sublattice $\Lambda_e$ of a
hypercubic lattice with orthogonal basis vectors $\hat a_I,
I=1,\dots,D$. We place the degrees of freedom on this even sublattice,
as well as operators $\Gamma^{(I,\alpha)}$ with $\alpha=1,\dots,n$
acting on these degrees of freedom. The operators
$\Gamma^{(I,\alpha)}$ are built as products of the $\gamma$-matrices
(in turn built from tensor products of Pauli matrices, see appendix
\ref{AA}). We take $\Gamma^{(I,1)}\equiv \gamma^I$, which we call
principal configuration. The need for the additional
$\Gamma^{(I,\alpha)}$ with $\alpha=2,\dots,n$ comes because the local
Hilbert is $2^n$ dimensional, and consequently $n$ operators are
necessary to gap the theory.

A generic $\Gamma$-operator can be parametrized in terms of a set of
integer-valued vectors $T_a^{(I,\alpha)}$, $a=1,\ldots,2n$, according
to
\begin{align}
\Gamma^{(I,\alpha)}
=
\left(\gamma^1\right)^{T^{(I,\alpha)}_1}\;
\left(\gamma^2\right)^{T^{(I,\alpha)}_2}\;
\dots\;
\left(\gamma^{2n}\right)^{T^{(I,\alpha)}_{2n}}.
\label{5.1}
\end{align}
Furthermore, since $(\gamma^I)^2=1$, only the values of the
$T$-vectors mod 2 matter. For the principal configuration we can
choose, for example, the vectors
\begin{align}
T^{(I,1)}_a \equiv t^{(I)}_a \equiv\delta^I_a,\;\; I=1,\dots,2n\;,
\quad {\rm and}\quad
T^{(2n+1,1)}_a\equiv t^{(2n+1)}_a\equiv -\sum_{I=1}^{2n}\;t^{(I)}_a
\;.
\label{5.2}
\end{align}
We call this choice as the canonical form. Written explicitly,
$t^{(1)}=(1,0,\dots,0)$, $t^{(2)}=(0,1,\ldots,0)$,$\dots$,
$t^{(2n)}=(0,0,\dots,1)$ and $t^{(2n+1)}=(-1,-1,\dots,-1)$.  The
condition $\sum_{I=1}^D t_a^{(I)}=0$ is tied to the
fact that all the $\gamma^I$ multiply to the identity (up to a
phase).

We define $\mathcal{O}^{(\alpha)}$ operators centered on the odd
sublattice $\Lambda_o$,
\begin{equation}
\label{eq1}
\mathcal{O}^{(\alpha)}_{\vec{x}}
\equiv
\prod^D_{I=1} \;\;
\Gamma^{(I,\alpha)}_{\vec{x}-\hat{a}_I}
\;\;
\Gamma^{(I,\alpha)}_{\vec{x}+\hat{a}_I}
\;,
\quad
\alpha=1,\dots,\frac{(D-1)}{2}.
\end{equation}
Notice that $\left(\mathcal{O}_{\vec{x}}^{(\alpha)}\right)^2=\openone$
follows because the $\Gamma^{(I,\alpha)}$ are products of Dirac
matrices. Using these operators we construct the Hamiltonian
\begin{equation}
\label{eq1b}
H=-\sum_{\alpha=1}^{(D-1)/2}
\;
\left(g_\alpha \;\sum_{\vec{x}}\mathcal{O}^{(\alpha)}_{\vec{x}} \right)
\;,
\end{equation}
where all coupling constants $g_{\alpha}$ are chosen to be
positive. We can further choose the operators $\Gamma^{(I,\alpha)}$
such that
\begin{equation}
\label{eq1c}
\left[
\mathcal{O}^{(\alpha)}_{\vec{x}}
\;,\;
\mathcal{O}^{(\beta)}_{\vec{x}'}
\right]
=
0\;,
\quad
\forall~ \alpha,\beta~~~\text{and}~~~ \forall~ {\vec{x}},{\vec{x}'}
\;.
\end{equation}
In this case, the Hamiltonian is a sum of commuting projectors and
there are as many commuting projectors (up to constraints that we
shall see in a moment give a topological degeneracy) as the number of
degrees of freedom in the problem.

\begin{figure}[!h]
	\centering
	\includegraphics[scale=0.7]{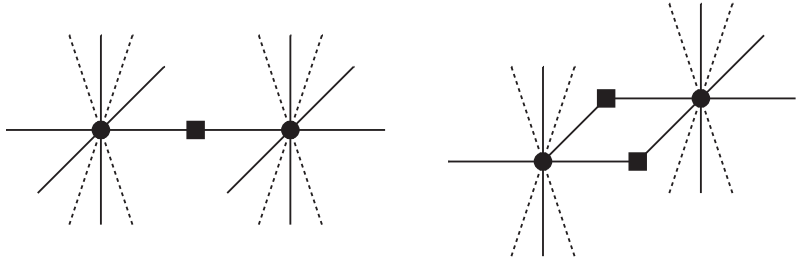}
	\caption{The two possibilities for distinct operators $\mathcal{O}$ sharing sites. The black squares correspond to the sites of the even sublattice $\Lambda_e$, while black dots correspond to the sites of the odd sublattice $\Lambda_o$. The dotted lines represent additional dimensions.}
	\label{lattice}
\end{figure}
The connection between the choice of operators $\Gamma^{(I,\alpha)}$
and the commutations between the $\mathcal{O}^{(\alpha)}_{\vec{x}}$
stems from the geometry imprinted via the definition Eq.~(\ref{eq1}),
and is depicted in Fig.~\ref{lattice}. The $\mathcal{O}^{(\alpha)}$'s
trivially commute both when they are defined at the same site $\vec
x\in\Lambda_o$ or when they do not share any sites (in $\Lambda_e$);
there just remains two cases to be checked: when they share one and
two sites. The neighboring $\mathcal{O}^{(\alpha)}$'s, defined at
sites $\vec{x}$ and $\vec{x}+2\hat{a}_I$ of $\Lambda_o$, share the
$\Lambda_e$ site at $\vec{x}+\hat{a}_I$, and they commute if 
\begin{align}
\label{eq:Gamma_condition1}
\left[
\Gamma^{(I,\alpha)}
\;,\;
\Gamma^{(I,\beta)}
\right]
=0
\;.
\end{align}
Neighboring $\mathcal{O}^{(\alpha)}$'s, defined at sites $\vec{x}$ and
$\vec{x}+\hat{a}_I+\hat{a}_J$ of $\Lambda_o$, share the two
$\Lambda_e$ sites at $\vec{x}+\hat{a}_I$ and $\vec{x}+\hat{a}_J$. The
operators on those sites either commute or anti-commute, which can be
cast as
\begin{align}
\Gamma^{(I,\alpha)}\;\;\Gamma^{(J,\beta)}
=
(-1)^{\eta^{(\alpha\beta)}_{IJ}}\;\;
\Gamma^{(J,\beta)}\;\;\Gamma^{(I,\alpha)}
\;,
\label{c1}
\end{align}
with $\eta^{(\alpha\beta)}_{IJ}=0$ or $1$. The desired commutation
relations Eq.~(\ref{eq1c}) are guaranteed if
\begin{align}
\label{eq:Gamma_condition2}
\eta^{(\alpha\beta)}_{IJ}
=
\eta^{(\alpha\beta)}_{JI}
\;.
\end{align}
In particular, the commutation (\ref{eq:Gamma_condition1}) implies
$\eta_{II}^{(\alpha,\beta)}=0$.

All these conditions can be satisfied using Dirac matrix
representations of the Clifford algebra. The simplest example is the
$D=3$ contained in Ref.~\cite{Chamon2005}, where one uses the $2\times 2$
representation
\begin{align}
\begin{matrix}
\Gamma^{(I,\alpha)} &\vline& I =& 1 & 2 & 3 & 
\\
\hline
\alpha=1&\vline& & \gamma^1 = \sigma^1\;\;& \gamma^2 = \sigma^2 \;\;&
\gamma^3 = \sigma^3 \;\;&
\end{matrix}
\end{align}
In $D=5$ we use the 4-dimensional representation of the Dirac matrices
and take the following $\Gamma^{(I,\alpha)}$ operators:
\begin{align}\label{d5}
\begin{matrix}
\Gamma^{(I,\alpha)} &\vline& I =& 1 & 2 & 3 & 4 & 5 &
\\
\hline
\alpha=1&\vline& & \gamma^1 & \gamma^2 & \gamma^3 & \gamma^4 & \gamma^5 &
\\
\alpha=2&\vline& &
\gamma^3 \gamma^5\; & \gamma^4 \gamma^5\; & \gamma^1 \gamma^5 \;&
\gamma^2 \gamma^5 \; & \gamma^5 &
\end{matrix}
\end{align}
that satisfy Eqs.~(\ref{eq:Gamma_condition1}) and (\ref{eq:Gamma_condition2}) and hence yield the set of commuting projectors $\mathcal{O}^{(1)}_{\vec{x}}$ and $\mathcal{O}^{(2)}_{\vec{x}}$, $\vec x\in\Lambda_o$. As we shall see in the next section, the field theory formulation provides a systematic way to construct operators of the sets $\alpha\geq2$ in arbitrary odd dimensions and satisfying all the required commutation rules.

The ground state of these models (for any $D=2n+1$) correspond to all
$\mathcal{O}^{(\alpha)}_{\vec{x}}$ having eigenvalue $+1$. Excitations
or defects correspond to those operators having instead eigenvalue
$-1$. That these models are fractonic requires that there is not a
single local operator whose net effect is to generate a defect pair or
equivalently move a single isolated defect. In $D=5$, for example, one
can easily check that there is no product of Dirac operators that
anti-commutes with a single $\Gamma^{(I,\alpha)}$ operator for given
$\alpha$. The minimum number of defects that can be created or
anihilated is four, like in the $D=3$ model of Ref.~\cite{Chamon2005}; this
number four remains the same for any odd dimension $D$ model. (See
appendix~\ref{AB} for details.)

A lower bound to the topological ground state degeneracy of the model
can be placed by noticing that the $\mathcal{O}^{(\alpha)}_{\vec{x}}$
have the property
\begin{equation}
\label{lat constraint}
\prod_{\vec{x}\in \Lambda_{o, k}}\mathcal{O}^{(\alpha)}_{\vec{x}} = \openone
\;,
\quad k=1,\ldots, 2^{D-1} ,\quad \alpha=1,\dots,(D-1)/2
\end{equation}
where $k$ accounts for all the $2^{D-1}$ sub-lattices $\Lambda_{o,
	k}$. Eq.\eqref{lat constraint} already give us a hint of the
degeneracy of the model, which will be at least
$2^{[(D-1)\;2^{D-2}]}$, but it happens that this number is a lower
bound to the degeneracy and in fact, the degeneracy can also depend on
the system size, as was shown in \cite{Terhal2011}. For a hypercube of volume $L^D$, the degeneracy dependence on the linear size $L$ is
$2^{(D-1)\;2^{D-3}\;L}$ (see appendix~\ref{AC} for details).


\section{Low-Energy Effective Field Theory}\label{effective theory}

In this section we shall derive an effective field theory capturing
the low-energy physical properties of the lattice models given by
(\ref{eq1b}). To connect the operators that act on the microscopic
degrees of freedom with the suitable operators that possess a
well-defined continuum limit, we define a map parametrized by the
vectors $T_a^{(I,\alpha)}$ of Eq. (\ref{5.1}):
\begin{equation}
\label{eq2}
\Gamma^{(I,\alpha)}_{\vec{x}} \equiv \exp\left(i \,T^{(I, \alpha)}_a \,K_{ab}\,A_b(\vec{x})\right)
\;, 
\end{equation}
where the repeated matrix indices $a,b=1,\ldots,2n$ are summed. We
need only $D-1=2n$ independent fields $A_a$ to construct all the
required operators. In addition, we see that this parametrization
introduces a symmetry
\begin{equation}
A\rightarrow Q\,A, \quad T^{(I, \alpha)}\rightarrow Q\,T^{(I, \alpha)}, \quad K\rightarrow \left(Q^\top\right)^{-1}\, K\,Q^{-1}
\;,
\label{eq2a}
\end{equation}
with $Q$ an arbitrary matrix.

For the case of the principal configuration, where
$\Gamma^{(I,1)}\equiv \gamma^I$ and, accordingly, $T_a^{(I,1)}\equiv
t^{(I)}_a$, we have
\begin{equation}
\label{princip}
\Gamma^{(I,1)}_{\vec{x}}\equiv\gamma^I_{\vec{x}}\equiv \exp\left(i\, t^{(I)}_a\, K_{ab}\,A_b(\vec{x})\right)\;.
\end{equation}
Properties of the fields $A_a$ and of the matrix $K$ can be obtained
from the analysis of the principal configuration. Indeed, we start by
computing
\begin{equation}
\label{eq4}
\gamma^I_{\vec{x}}\;\gamma^J_{\vec{x}^{\prime}}
=
\exp\left(-\left[t^{(I)}_a\, K_{ab}\, A_b(\vec{x})\;,\; t^{ (J)}_{a^\prime}\,K_{a^\prime b^\prime}\,A_{b^{\prime}}(\vec{x}')\right]\right)
\;\gamma^J_{\vec{x}^{\prime}}\;\gamma^I_{\vec{x}}
\;,
\end{equation}
where we have used the BCH formula and assumed that
the commutator appearing in this expression is a $c$-number. Since
$\gamma^I_{\vec{x}}$ and $ \gamma^J_{\vec{x}^{\prime}}$ must commute
if $\vec{x}\neq\vec{x}'$ and anti-commute if $\vec{x}=\vec{x}'$ and
$I\neq J$, we impose
\begin{equation}
\label{eq5}
\left[A_b\left(\vec{x}\right),
A_{b^{\prime}}\left(\vec{x}^{\prime}\right)\right]
\equiv
i\pi\,
(K^{-1})_{bb^{\prime}}\;\delta_{\vec{x}\vec{x}^\prime}
\;,
\end{equation}
which will be interpreted as an equal-time commutation relation in a
field theory formulation with a canonical pair $A_a(\vec{x})$ and
$\Pi_a(\vec{x})=\frac{1}{\pi}\;(K^T)_{ab}\;A_b$. We shall return to
this point later. Using the commutation relation (\ref{eq5}) in
(\ref{eq4}) leads to the further following conditions to match the
anti-commutation relations among the $\gamma^I_{\vec{x}}$:
\begin{equation}
\label{eq6}
t^{(I)}_a \;(K^{\top})_{ab} \;\,t^{(J)}_b =
\begin{cases}
\;1\!\!\mod 2, &\!  ~ I\neq J \\
\;0\!\!\mod 2, &\!  ~ I=J\;.
\end{cases}
\end{equation}
The condition \eqref{eq6} is particular to the principal
configuration, and reflects that the building blocks of the theory are
anti-commuting objects (Dirac matrices); it is not needed for the
other $T$-vectors with $\alpha\geq 2$, since the associated operators
(products of Dirac matrices) may either commute or anti-commute. Notice that the conditions in \eqref{eq6} imply that the fields are compact, since the shifts
\begin{equation}
A_b\rightarrow A_b +2 \pi \sum_{J=1}^{D}t_b^{(J)} m_J\,,~~~m_J\in\mathbb{Z}\,,
\end{equation}
do not change $\gamma^I_{\vec{x}}$ because
\begin{equation}
\exp \left(2 i \pi \sum_{J=1}^D t_a^{(I)}K_{ab} t_b^{(J)} m_J  \right)=1\,. 
\end{equation}

The most general solution (independent of $t$) for the condition in
the second line of \eqref{eq6} corresponds to the case where $K$ is an
anti-symmetric matrix. In writing the commutation relation (\ref{eq5})
we assumed that the inverse $K^{-1}$ exists, which requires $\det
K\neq 0$, a condition only possible to satisfy for even-dimensional
anti-symmetric $K$ matrices. Recall that $K$ is a $2n\times
2n=D-1\times D-1$ matrix, so the construction works for odd dimensions
$D$.

A useful relation involving the $t$-vectors emerges when we consider
the product of all matrices $\gamma$ in the same site. Suppressing the
matrix indices for simplicity, we have
\begin{eqnarray}
\openone\sim\gamma^1 \gamma^2\cdots \gamma^D
&=&e^{it^{(1)}\,K\,A}e^{it^{(2)}\,K\,A}\cdots e^{it^{(D)}\,K\,A}\nonumber\\
&=&\exp\left[ i \sum_{I=1}^{D}t^{(I)}\,K\,A\right]\;\exp\left[\frac{\pi i}{2} \sum_{I<J}t^{(I)}\,K\,t^{(J)}\right]\nonumber\\
&=&\exp\left[ i \sum_{I=1}^{D}t^{(I)}\,K\,A\right]\;\exp\left[\frac{\pi i}{2}\, \frac{(D-1)\,D}{2}\right],
\label{eq6a}
\end{eqnarray}
where we have used (\ref{eq6}) in the last step. In order for the
right-hand side to be proportional to the identity, we require
\begin{equation}
\sum_{I=1}^{D}t^{(I)}_a=0
\;,
\label{eq6b}
\end{equation} 
which we refer to as the neutrality condition. Notice that this is satisfied with the choice in (\ref{5.2}). Moreover, we shall require the neutrality condition for all sets of operators $\Gamma^{(I,\alpha)}$, which corresponds to 
\begin{equation}
\label{eq13}
\sum_{I=1}^D T^{(I, \alpha)}_a=0
\;.
\end{equation}

We now proceed by analyzing the field theory counterpart of the
lattice operators (\ref{eq1}). Using the representation \eqref{eq2},
it follows that
\begin{equation}
\label{eq8}
\mathcal{O}^{(\alpha)}_{\vec{x}}
=
\exp \left[i \sum^{D} _{J=1} \left(\;T^{(J, \alpha)}_a \,K_{ab} \,A_b(\vec{x}+\hat{a}_J)
+
T^{ (J, \alpha)}_{a} \,K_{ab} \,A_{b} (\vec{x}-\hat{a}_J)\;\right)\right] \;.
\end{equation}
Given the commutation relation (\ref{eq5}), we have to determine the
conditions on $T^{(J,\alpha)}$ and $K$ so as to produce commuting
operators $\mathcal{O}^{(\alpha)}_{\vec{x}}$, i.e., so that the field
theory representation on the right-hand side reproduces the
commutations in Eq. (\ref{eq1c}). As discussed in the previous
section, there are two situations where a nontrivial commutation rule
may arise: when the operators share one or two sites. When they do not
share any site, the commutation is trivially satisfied. To take into
account these two situations, we just have to consider two operators
$\mathcal{O}^{(\alpha)}_{\vec{x}}$ at positions $\vec{x}$ and
$\vec{x}+\vec{a}_I+\vec{a}_J$. Thus, if $I=J$ the operators share the
site $\vec{x}+\vec{a}_I$ and if $I\neq J$ they share the two sites
$\vec{x}+\vec{a}_I$ and $\vec{x}+\vec{a}_J$. The requirement of
commutation between the operators $\mathcal{O}^{(\alpha)}_{\vec{x}}$
and $\mathcal{O}^{(\beta)}_{\vec{x}+\vec{a}_I+\vec{a}_J}$ is
\begin{equation}
\label{eq13a}
C^{(\alpha\beta)}_ {IJ}=0
\;,
\end{equation}
where 
\begin{equation}
C^{(\alpha\beta)}_ {IJ}\equiv T^{(I,\alpha)}_a\, K_{ab}\, T^{(J,\beta)}_b + T^{(J,\alpha)}_a \,K_{ab}\, T^{(I,\beta)}_b\;.
\label{eq13ab}
\end{equation}
Notice that $C^{(\alpha\beta)}_ {IJ}=C^{(\alpha\beta)}_ {JI}$ and
$C^{(\alpha\beta)}_ {IJ}=-C^{(\beta\alpha)}_ {IJ}$. The symmetry in
the $IJ$ indices follows directly from the way $C^{(\alpha\beta)}_
{IJ}$ is defined, whereas the anti-symmetry in the $\alpha\beta$
indices follows from the anti-symmetry of the matrix $K$. In
particular, the condition (\ref{eq13a}) is automatically satisfied if
$\alpha=\beta$, which is consistent with the fact that
$\mathcal{O}^{(\alpha)}_{\vec{x}}$ operators of the same kind commute
with each other. A systematic procedure for constructing $T$-vectors
satisfying the condition (\ref{eq13a}) is presented in appendix
\ref{AB}.
Next we consider the continuum limit of the relation
\eqref{eq8}. The expansion of the field $A_b$
reads
\begin{equation}
A_{b}(\vec{x}\pm \hat{a}_J)
=
A_{b}(\vec{x})\pm \sum_I \;a_J^I \;\partial_I \;A_b(\vec{x})
+\frac{1}{2}\sum_{I,K}a_J^I\, a_J^K\;\partial_I\,\partial_K \;A_b(\vec{x})+\cdots\;.
\end{equation}
As the unit vectors $\hat{a}_J$ have the components $a_J^I=\delta_J^I$, we get 
\begin{align}
\label{eq9}
\mathcal{O}^{(\alpha)}_{\vec{x}}
=
\exp \left(2i\sum_{J=1}^{D}T^{(J,\,\alpha)}_a\,K_{ab}\, A_{b}(\vec{x})
+
i\,\sum_{J=1}^{D}T^{(J,\,\alpha)}_a\,K_{ab}\,\partial^2_{J} \,A_{b}(\vec{x})+\ldots\right)\;.
\end{align}
We see that the neutrality condition \eqref{eq13} ensures that the first term in the exponential vanishes, so that the operator $\mathcal{O}^{(\alpha)}_{\vec{x}}$ reduces to
\begin{align}
\label{eq9a}
\mathcal{O}^{(\alpha)}_{\vec{x}}
=
\exp \left(i\sum_{J=1}^{D}T^{(J,\,\alpha)}_a\,K_{ab}\,\partial^2_{J}\, A_{b}(\vec{x})+\cdots\right)
\;.
\end{align}
The Hamiltonian in (\ref{eq1b}) becomes
\begin{equation}
H\sim-2\sum_{\alpha}g_{\alpha} \int d^Dx\; \cos \left(M^{(\alpha)}(\vec{x})\right)
\;,
\label{eq9b}
\end{equation}
with 
\begin{equation}
M^{(\alpha)}(\vec{x})\equiv\sum_{J=1}^{D} T^{(J,\,\alpha)}_a\, K_{ab}\,\partial^2_{J}\, A_{b}(\vec{x})
\;.
\label{eq9b1}
\end{equation}
We see that the ground state corresponds to the case where all the
cosines in \eqref{eq9b} are simultaneously pinned at
$M^{(\alpha)}=2\pi m^{(\alpha)}$ for all sites, where
$m^{(\alpha)}\in\mathbb{Z}$. We can enforce this in a corresponding
field theory description of the ground state through a Lagrange
multiplier, as we will discuss in a moment.

Before going to the field theory it is convenient to express the
operator $M^{(\alpha)}(\vec{x})$ in a way that solves the constraint
of the neutrality condition \eqref{eq13}. Thus, we single out one of
the directions, say the last one $J=D$, and write
\begin{eqnarray}
M^{(\alpha)}(\vec{x})
&=&
\sum_{J=1}^{D-1}T^{(J,\,\alpha)}_a K_{ab}\,\partial^2_{J}\, A_{b}(\vec{x})
\;+\;
T^{(D,\,\alpha)}_a\, K_{ab}\, \partial^2_{D} \, A_{b}(\vec{x})\nonumber\\
&=&
\sum_{J=1}^{D-1}T^{(J,\,\alpha)}_a \, K_{ab} \, D_{J}\, A_{b}(\vec{x})
\;,
\label{eq9c}
\end{eqnarray}
where the derivative operator $D_J$ is defined as $D_{J}\equiv
\partial_{J}^2-\partial_D^2$. It is also convenient to define another
differential operator as
\begin{eqnarray}
\mathcal{D}^{(\alpha)}_a
&\equiv& \sum_{J=1}^{D} T^{(J,\,\alpha)}_a\, \partial^2_{J}\nonumber\\
&=&
\sum_{J=1}^{D-1} T^{(J,\,\alpha)}_a \, D_{J}
\;.
\label{eq9d}
\end{eqnarray}
In terms of $\mathcal{D}^{(\alpha)}_a$, the operator
$M^{(\alpha)}(\vec{x})$ in \eqref{eq9c} acquires a simple compact form
\begin{equation}
\label{eq15}
M^{(\alpha)}(\vec{x})= K_{ab}\,\mathcal{D}^{(\alpha)}_a\,A_b\;,
\end{equation}
which makes evident its invariance under gauge transformations
\begin{equation}
\label{eq16}
A_a \rightarrow A_a + \sum_{\alpha}\mathcal{D}^{(\alpha)}_a \zeta^{(\alpha)}
\;,
\end{equation}
with $\zeta^{(\alpha)}=\zeta^{(\alpha)}\left(\vec{x}, t\right)$ being
a set of arbitrary functions of spacetime coordinates. In fact, notice
that
\begin{align}
K_{ab}\;\mathcal{D}^{(\alpha)}_a\;\mathcal{D}^{(\beta)}_b\;
&=
\sum_{I,J=1}^D\;
K_{ab}\;T^{(I,\,\alpha)}_a\; T^{ (J,\,\beta)}_{b}
\;\partial^2_I\;\partial^2_J
\nonumber\\
&=
\sum_{I,J=1}^D\;C^{(\alpha\beta)}_{IJ}
\;\partial^2_I\;\partial^2_J
\nonumber\\
&=0\;,\quad{\rm since}\;C^{(\alpha\beta)}_{IJ}=0
\;.
\end{align}
Therefore, the condition above, needed for gauge invariance, is
precisely the condition for commutation of the cosine operators
\eqref{eq13a}.

With all these elements in place, we can write down a field theory
which describes the ground state of the microscopic fracton model,
\begin{equation}
\label{eq18}
S
=
\int d^Dx\; dt\;
\frac{1}{2\pi}\left[ K_{ab}\; A_a\; \partial_0 \; A_b
+
2\sum_{\alpha} A^{(\alpha)}_0\,K_{ab}\;\mathcal{D}^{(\alpha)}_a\;A_b\right]\;.
\end{equation}
The first term is responsible for the commutation relation
\eqref{eq5}\footnote{Notice that the prefactor of $\frac{1}{2}$ in the
	action (\ref{eq18}) ensures the right numerical factor in the
	commutation relation (\ref{eq5}), since for each pair of coordinates
	we always have two contributions because of the anti-symmetry of the
	matrix $K$, for example, $K_{12}\,(A_1\partial_0 A_2-A_2\partial_0
	A_1)$. This pair of terms must be brought into a single term through
	integration by parts before computing the canonical momentum.},
whereas the second one enforces the ground state constraints, with
$A^{(\alpha)}_0$ a set of Lagrange multipliers. The requirement of
full gauge invariance of the action (up to boundary terms) dictates
that $A^{(\alpha)}_0$ must transform as
\begin{equation}
\label{eq21}
A^{(\alpha)}_0 \rightarrow A^{(\alpha)}_0 +\partial_0 \zeta^{(\alpha)}.
\end{equation}
Thus, we end up with a {\it bona fide} gauge theory, which resembles
the Chern-Simons description of topologically ordered systems. The
gauge-invariant ``electric'' and ``magnetic'' fields can be defined as
\begin{equation}
E_a
\equiv
\partial_0 A_a - \sum_{\alpha}\mathcal{D}^{(\alpha)}_a\, A^{(\alpha)}_0
\quad\text{and}\quad
B^{(\alpha)}_{a_1 a_2 \cdots a_{D-3}}\equiv\epsilon_{a_1 a_2 \cdots a_{D-1}}
\;\mathcal{D}^{(\alpha)}_{a_{D-2}}\,A_{a_{D-1}},
\label{ebgauge}
\end{equation}
where $\epsilon_{a_1 a_2 \cdots a_{D-1}}$ is the Levi-Civita tensor of rank $D-1$. 

\section{Properties of the Effective Theory}\label{Properties}

\subsection{Level Quantization}\label{PropertiesA}

Now we will explore some properties of the effective field theory
(\ref{eq18}). Firstly, it is interesting to understand whether there
is a notion of quantization of the ``level'' of the theory, which in the
present case is given by the matrix $K$. To address this question we
consider the principal configuration $T^{(I,1)}=t^{(I)}$. In this
case, the $t$-vectors must satisfy the conditions in
(\ref{eq6}). Then, we use the symmetry transformations in (\ref{eq2a})
to make a specific choice for the $t$-vectors. For example, if we pick
up the canonical form (\ref{5.2}), we obtain the following level
quantization condition:
\begin{equation}
K_{IJ}=\text{odd}\,,~~~\text{with}~I\neq J~~~\text{and}~~~   I,J=1,\ldots,D-1\;,
\label{eq21b}
\end{equation}
i.e., all the off-diagonal elements must be odd integers, and
consequently nonvanishing. Of course, different representations of the
$t$-vectors yield different quantization of the elements of the
matrix $K$, but in all the cases we end up with some notion of
quantization due to the conditions in (\ref{eq6}).

{From the field theory alone, the quantization of the level can be understood as follows. Consider a manifold $M=\mathcal{S}^1\times\,\mathcal{M}^D$, with $\mathcal{S}^1$ representing the time direction with period $[0,\,\tau)$ and $\mathcal{M}^D$ a spatially closed manifold. Due to the compact nature of the fields $A_a$, we have a quantized flux
\begin{equation}
\int_{\mathcal{M}^D} B^{(\alpha)}_{a_1a_2\ldots a_{D-3}}\equiv \pi\, p^{(\alpha)}_{a_1a_2\ldots a_{D-3}}, ~~~p^{(\alpha)}_{a_1a_2\ldots a_{D-3}}\in \mathbb{Z}\,.
\end{equation}
Consider large gauge transformations that wind around the time direction. The $A_0^{(\alpha)}$ field transform as
\begin{equation}
A_0^{(\alpha)} \rightarrow A_0^{(\alpha)} + \frac{2\pi}{\tau}\,n^{(\alpha)}\,,~~~n^{(\alpha)}\in \mathbb{Z}\,,
\label{large}
\end{equation}
and the corresponding variation of the action under these transformations is
\begin{align}
\delta S &= \pi\,K_{ab} \sum_{\alpha}\frac{1}{(D-3)!}\; n^{(\alpha)}\, \epsilon_{a_1a_2\ldots a_{D-3}a b}\; p^{(\alpha)}_{a_1a_2\ldots a_{D-3}}\,,\nonumber\\
&=\pi\,K_{ab}\,\mathbb{Z}_{ab}\,
\end{align}
where $\mathbb{Z}_{ab}$ an integer valued quantity obtained from the summation above. From this, it is straightforward to note that in order for the quantum theory to be invariant under the large gauge transformations \eqref{large}, the $K$-matrix elements have to be integer valued. Further details of this calculation are found in appendix \ref{AF}
}
\subsection{Three dimensional case}	

It is instructive to compare the effective field theory that we have
obtained for the particular case of three spatial dimensions, $D=3$, with the result of
Ref.~\cite{Burnell2019}. For $D=3$ there is only one configuration, the
principal configuration $\alpha=1$. The matrix $K$ in this case is
\begin{equation}
K=\left(
\begin{array}{cc}
0& k\\
-k& 0\\
\end{array}
\right)
~~~\text{and}~~~
K^{-1}=\left(
\begin{array}{cc}
0& -\frac{1}{k}\\
\frac{1}{k}& 0\\
\end{array}
\right).
\end{equation}
The action, in terms of electric and magnetic fields, reduces to 
\begin{equation}
\label{eq18d}
S=\int d^3x\, dt  \frac{k}{2\pi}\,\left[A_1\, E_2-A_2\, E_1+A_0 \,B \right],~~~[A_1(\vec{x}),\,A_2(\vec{x}')]=-\frac{\pi i}{k}\,\delta\left(\vec{x}-\vec{x}'\right)\;.
\end{equation} 
With the canonical choice (\ref{5.2}), the derivative operators
entering the electric and magnetic fields become
$\mathcal{D}_1=\partial_1^2-\partial_3^2$ and
$\mathcal{D}_2=\partial_2^2-\partial_3^2$, whereas the coefficient $k$
must be an odd integer, in accordance with (\ref{eq21b}). In order to
compare with \cite{Burnell2019}, we just need to rename the fields and the
derivative operators according to $A_1\rightarrow-A_2$,
$A_2\rightarrow A_1$, $\mathcal{D}_1\rightarrow-\mathcal{D}_2$, and
$\mathcal{D}_2\rightarrow \mathcal{D}_1$ (see equation (95) of
\cite{Burnell2019}), which leave both the action and the commutation relation
in (\ref{eq18d}) unchanged. In this form, we can immediately compare
with the results of \cite{Burnell2019} with the following identification
between the parameters $k=s/2$. In that work, the original Chamon
model (with full cubic symmetry) is recovered for $s=2$ (in \cite{Burnell2019},
the level quantization is $s\in\mathbb{Z}$), which in our
normalization corresponds to $k=1$. This choice describes the
$2$-state system at each site, as expected. Also, this choice of $k$
is allowed by the level quantization (\ref{eq21b}) associated with the
canonical choice for the $t$-vectors. {For a general discussion of how the $K$-matrix elements are determined by the microscopic theory, we refer the reader to appendix \ref{AE}}.
\subsection{Conservation Laws}\label{charge conservation}

A gauge-invariant coupling to matter can be introduced in the action \eqref{eq18} through the terms $\sum_{\alpha}A^{(\alpha)}_0J^{(\alpha)}_0+A_aJ_a$, provided that the current satisfies the continuity equation
\begin{equation}
\label{continuity}
\partial^{\; }_0\, J^{(\alpha)}_0 = \mathcal{D}^{(\alpha)}_a \,J_a\;.
\end{equation}
By integrating over the whole space and assuming periodic boundary
conditions along all directions, it follows that charge is conserved
in the whole system,
\begin{equation}
\label{charge}
\frac{d}{dt}\int d^D x \;J^{(\alpha)}_0 = \int d^D x\; \mathcal{D}^{(\alpha)}_a\, J_a=0\;
\end{equation}
In addition, given the form of the derivative operators
$\mathcal{D}^{(\alpha)}_a$, we also have more restrictive conservation
laws. These extra conservation laws require that charge is also
conserved on a set of sub-manifolds of the system. It is due to these
extra conservation laws that the fracton behavior of the excitations
emerges.

To find the sub-manifolds where charge is conserved, we use the
definition of the derivative operators $\mathcal{D}_a^{(\alpha)}$ in
\eqref{eq9d} to write the continuity equation as
\begin{eqnarray}
\label{eq36}
\partial^{\;}_0\, J^{(\alpha)}_0
&=&\sum_{I=1}^{D-1} T^{(I,\alpha)}_a\; D_I\;J^{\;}_a\,,\nonumber\\
&=&\sum_{I=1}^{D-1} D_I\; J_{I}^{(\alpha)},
\end{eqnarray}
where we have defined $J_{I}^{(\alpha)}\equiv T^{(I,\alpha)}_a
J_a$. Recalling that $D_I\equiv \partial_I^2-\partial_D^2$, it is
convenient to introduced the directions
$\hat{x}^{\sigma_I}_{ID}\equiv\hat{a}_I+\sigma_I\,
\hat{a}_D$, with $\sigma_I=\pm 1$. In this notation,
\eqref{eq36} can be written as
\begin{equation}
\label{eq37}
\partial^{\;}_0\, J^{(\alpha)}_0
=
4\sum_{I=1}^{D-1}\left(\partial^-_{{ID}}\;\partial^+_{{ID}}\right) J_{I}^{(\alpha)}\; .
\end{equation}
This form of the continuity equation induces $2^{D-1}$ extra
conservation laws, explicitly, that charge must be conserved in each
of the $(D-1)$-dimensional sub-manifolds labelled by
$\left(x^{\sigma_1}_{1D}, \dots, x^{\sigma_{D-1}}_{(D-1)D}\right)$. Indeed,
if we integrate $J^{(\alpha)}_0$ over any of these sub-dimensional
manifolds, we obtain the following conserved charges
\begin{equation}
Q^{(\alpha)}_{(\sigma_1,\,\sigma_2,\,\dots,\, \sigma_{D-1})}
\equiv
\int dx_{1D}^{\sigma_1}\, dx_{2D}^{\sigma_2}\,\dots\, dx_{(D-1)D}^{\sigma_{D-1}} \;\, J_0^{(\alpha)}\;.
\label{charge2}
\end{equation}
These conservation laws, in turn, imply that the dipole moment in the
direction perpendicular to those manifolds is conserved. Naturally,
such conservation laws impose several restrictions on the mobility of
the particles. We build in detail the form of the excitations in appendix \ref{AD}.
\subsection{Ground State Degeneracy}\label{GSD}

Here we discuss the computation of the ground state degeneracy using the effective field theory. Naturally, in the continuum limit the degeneracy is infinite so that we shall adopt some kind of discretization (regularization) of the theory. The form of the conservation laws in the sub-dimensional manifolds provides a very natural way to discretize the theory in a layered structure. The basic idea is to consider the system as a stack of layers corresponding to the sub-dimensional manifolds where charge is conserved. 

Let us start with the case $D=3$. The action (\ref{eq18}) reduces to
\begin{equation}
S=\int d^3x\, dt \, \frac{k}{\pi}\, A_1\, \partial_0\, A_2 + \cdots,
\label{eq38}
\end{equation}
where we keep explicitly only the part relevant for the computation of the degeneracy. In this case, charge is conserved in $2^2$ sub-spaces labeled by $\sigma_1,\sigma_2=\pm$, with the corresponding measures 
\begin{equation}
\int dx_{13}^{\sigma_1}\,dx_{23}^{\sigma_2} \,.
\label{eq39}
\end{equation}
The strategy is to write the action (\ref{eq38}) in terms of the coordinates $x_{13}^{\sigma_1}\,,x_{23}^{\sigma_2}\,,x_{\perp}$, where $x_{\perp}$ is the coordinate perpendicular to the plane defined by the directions $x_{13}^{\sigma_1}$ and $x_{23}^{\sigma_2}$. Upon this change of variables,
\begin{equation}
\int d^3x\rightarrow \int dx_{13}^{\sigma_1}\, dx_{23}^{\sigma_2}\, dx_{\perp}\, \mathcal{J}\,,
\label{eq40}
\end{equation}
where $\mathcal{J}$ is the Jacobian of the transformation. As this transformation is linear, $\mathcal{J}$ is just a constant and can be absorbed in $d x_{\perp}$. The transformation from the coordinates $x_1,\,x_2,\,x_3$ to $x_{13}^{\sigma_1},\,x_{23}^{\sigma_2},\,x_{\perp}$ will change the limits of integration. However, as the ground state degeneracy in each plane with periodic boundary conditions (forming a torus $T^2$) does not depend on the area of the plane (torus), so we can ignore the area of integration in our computation as long as we assume periodic boundary conditions along the plane $x_{13}^{\sigma_1}$-$x_{23}^{\sigma_2}$.

The next step it to discretize the coordinate $x_{\perp}$. We consider that the perpendicular direction is composed by a stack of $N$ layers, 
\begin{equation}
\int dx_{\perp} \rightarrow \sum_{i=1}^{N} \,2a\,,
\label{eq41} 
\end{equation}
where $2a$ is the separation between the planes, twice the lattice spacing of the microscopic model. This discretization ties the number of layers to the linear size: $N=L/2a$. (Equivalently $N=L/2$ given we set $a=1$). The gauge fields $A_a$ need to be rescaled properly
\begin{equation}
A_a(t,x_{13}^{\sigma_1},x_{23}^{\sigma_2},x_{\perp})\rightarrow \frac{1}{\sqrt{2a}}\, A_a^i(t,x_{13}^{\sigma_1},x_{23}^{\sigma_2})\,.
\label{eq42}
\end{equation}
The action (\ref{eq38}) becomes
\begin{equation}
S=\sum_{i=1}^{N}\, \int dt\, dx_{13}^{\sigma_1}\, dx_{23}^{\sigma_2}\,  \frac{k}{\pi}\, A_1^i\, \partial_0\, A_2^i + \cdots\,.
\label{eq43}
\end{equation}
Thus we end up with $N$ copies of (2+1)-dimensional theories. 

The dimension of the gauge fields in mass units is $[A_a]=D/2$. After discretization, the rescaled fields in (\ref{eq42}) have dimension $[A_a^i]=\frac{D-1}{2}$. In particular, $[A_a^i]=1$ for $D=3$. Therefore, for each of the layers, we can define the holonomies 
\begin{equation}
\exp  \left(i\,\int_0^{l_i} dx_{13}^{\sigma_1}\, A_1^i\right)~~~\text{and}~~~ \exp  \left(i\,\int_0^{l_i} dx_{23}^{\sigma_2}\, A_2^i\right)\,,
\label{eq44}
\end{equation}
where $l_i$ is the size of each cycle of the 2-torus, and the arguments of the exponentials are properly dimensionless. These objects are gauge-invariant. In fact, under a gauge transformation, the fields transform as 
\begin{equation}
A_1^i\rightarrow A_1^i\, +\,\partial_{13}^+\,\partial_{13}^-\,\zeta^i~~~\text{and}~~~A_2^i\rightarrow A_2^i\, +\,\partial_{23}^+\,\partial_{23}^-\,\zeta^i\,.
\label{eq45}
\end{equation}
Let us analyse, say, the first holonomy in (\ref{eq44}). Under a gauge transformation, it changes by a factor 
\begin{equation}
\exp \left(i \,\int_0^{l_i} dx_{13}^{\sigma_1}\, \partial_{13}^+\,\partial_{13}^-\,\zeta^i\right)\, =\, \exp\left( i \, \partial_{13}^{-\sigma_1}\,\zeta^i\,\Big{|}_{x_{13}^{\sigma_1}=0}^{x_{13}^{\sigma_1}=l_i}\right)\, \equiv 1\,.
\label{eq46}
\end{equation}
The above condition is satisfied with the general periodic boundary condition
\begin{equation}
\zeta^i \big{|}_{x_{13}^{\sigma_1}=l_i}-\zeta^i\big{|}_{x_{13}^{\sigma_1}=0}=2\pi\, n_1^i\, x_{13}^{-\sigma_1},~~~n_1^i\in \mathbb{Z}\,.
\label{eq47}
\end{equation}
Infinitesimal gauge transformations correspond to $n_1^i=0$, whereas $n_1^i\neq 0$ are associated with large gauge transformations. Similarly, for the second holonomy in (\ref{eq44}), we obtain
\begin{equation}
\zeta^i\, \big{|}_{x_{23}^{\sigma_2}=l_i}-\zeta^i\,\big{|}_{x_{23}^{\sigma_2}=0}=2\pi\, n_2^i\, x_{23}^{-\sigma_2}\,,~~~n_2^i\in \mathbb{Z}\,.
\label{eq48}
\end{equation}
A large gauge transformation satisfying all these conditions can be constructed explicitly, 
\begin{equation}
\zeta^i = \frac{2\pi\, n_1^i}{l_i}\, x_{13}^+\, x_{13}^-\,+ \frac{2\pi \,n_2^i}{l_i}\, x_{23}^+\, x_{23}^-,~~~n_1^i,\, n_2^i\in \mathbb{Z}\,.
\label{eq49}
\end{equation}
This implies an equivalence for the gauge fields
\begin{equation}
A_1^i\cong A_1^i+ \frac{2\pi}{l_i}\,m_1^i~~~\text{and}~~~A_2^i\cong A_2^i+ \frac{2\pi}{l_i}\, m_2^i, ~~~m_1^i, \,m_2^i\in \mathbb{Z}\,.
\label{eq50}
\end{equation}

Now we consider the ground state configuration, which corresponds to solutions depending only on the time,
\begin{equation}
A_a^i(t,x_{13}^+,x_{23}^-)=\frac{1}{l_i}\,\bar{A}_a^i(t)\,.
\label{eq51}
\end{equation}
Plugging this equation into the action (\ref{eq43}) we obtain
\begin{equation}
S=\sum_{i=1}^{N}\, \int dt\,  \frac{k}{\pi}\, \bar{A}_1^i\, \partial_0\, \bar{A}_2^i\,.  
\label{eq52}
\end{equation}
The holonomies become
\begin{equation}
e^{i \bar{A}_1^i}~~~\text{and}~~~ e^{i\bar{A}_2^i}\,.
\label{eq53}
\end{equation}
From the action (\ref{eq51}) it follows the commutation rule
\begin{equation}
[\bar{A}_1^i,\,\bar{A}_2^j]=-\frac{i\pi }{k}\,\delta^{ij}\,,
\label{eq54}
\end{equation}
leading to the commutation relation between the holonomies
\begin{equation}
e^{i\bar{A}_1^i}\;e^{i\bar{A}_2^i}=e^{i\bar{A}_2^i}\;e^{i\bar{A}_1^i}\;e^{\frac{i\pi}{k}}\,,
\label{eq55}
\end{equation}
which implies a $2k$-fold degeneracy for each plane $i$. The degeneracy of the layered system is then 
\begin{equation}
(2k)^{N}\,.
\label{eq56}
\end{equation}
Finally, taking into account that we have 4 sub-dimensional manifolds where charge is conserved, the total degeneracy is
\begin{equation}
\text{GSD}=(2k)^{4N}\,.
\label{eq57}
\end{equation}
Using that $N= L/2$, we recover the degeneracy of the lattice model in $D=3$. For $k=1$ it agrees with the result of \cite{Terhal2011}: $2^{2L}$.  

Now let us discuss how this generalizes to higher dimensional spaces. For concreteness, we consider the (5+1)-dimensional action
\begin{equation}
S=\int dt\, d^5x\, \frac{1}{2\pi}\,K_{ab}\,A_a\,\partial_0\, A_b+\cdots\,.
\label{eq58}
\end{equation}
In this case, charge is conserved in the following $2^4$ sub-dimensional manifolds with the corresponding measures,
\begin{equation}
\int  dx_{15}^{\sigma_1}\, dx_{25}^{\sigma_2}\, dx_{35}^{\sigma_3}\, dx_{45}^{\sigma_4}\,.
\label{eq59}
\end{equation}
We proceed similarly to the previous case, i.e., we write the action in terms of the coordinates of a sub-manifold where charge is conserved plus a perpendicular direction $x_{\perp}$, which is then discretized. With this, the action (\ref{eq58}) becomes 
\begin{equation}
S=\sum_{i=1}^N\int dt\, dx_{15}^{\sigma_1}\, dx_{25}^{\sigma_2}\, dx_{35}^{\sigma_3}\, dx_{45}^{\sigma_4}\,  \frac{1}{2\pi}\,K_{ab}\,A_a^i\,\partial_0\, A_b^i+\cdots\,,
\label{eq60}
\end{equation}
where the fields $A_a^i$ were rescaled as in (\ref{eq42}). 

Now, the key point is that we can rotate the matrix $K$ according to (\ref{eq2a}) to bring it to the block-diagonal form
\begin{equation}
Q\,K\,Q^{T}=\text{Diag}\left\{
\left(
\begin{array}{cc}
0& k_1\\
-k_1& 0\\
\end{array}
\right),
\left(
\begin{array}{cc}
0& k_2\\
-k_2& 0\\
\end{array}
\right)
\right\},
\label{eq61}
\end{equation}
where $k_1$ and $k_2$ are real and positive. In this basis, the fields $A_a^i$ decouple pairwise, 
\begin{equation}
S=\sum_{i=1}^N\int dt\, dx_{15}^{\sigma_1}\, dx_{25}^{\sigma_2}\, dx_{35}^{\sigma_3}\, dx_{45}^{\sigma_4}\,\left[  \frac{k_1}{\pi}\,A_1^i\,\partial_0\, A_2^i+ \frac{k_2}{\pi}\,A_3^i\,\partial_0\, A_4^i+\cdots\right]\,.
\label{eq62}
\end{equation}
Thus, we can construct the following pairs of holonomies
\begin{equation}
\exp \left(i \,l_i \int_0^{l_i} dx_{15}^{\sigma_1}\, A_1^i\right)~~~\text{and}~~~ \exp \left(i\, l_i\int_0^{l_i} dx_{25}^{\sigma_2}\, A_2^i\right)\,,
\label{eq63}
\end{equation}
and
\begin{equation}
\exp \left(i \,l_i\int_0^{l_i} dx_{35}^{\sigma_3}\, A_3^i\right)~~~\text{and}~~~\exp \left(i\, l_i \int_0^{l_i} dx_{45}^{\sigma_4}\, A_4^i\right) \,.
\label{eq64}
\end{equation}
Notice that we have introduced an appropriate factor of $l_i$ in order to have a dimensionless argument in the exponentials\footnote{In an arbitrary odd $D$-dimensional space, as $[A_a^1]=\frac{D-1}{2}$, we shall include the factor $l_i^{\frac{D-3}{2}}$ in order to make the argument dimensionless, i.e., the holonomies are of the form: $\exp \left(i\, l_i^{(D-3)/2} \int_0^{l_i} dx_{aD}^{\sigma_a}\, A_a^i\right)$, with $a=1,2,\ldots D-1$.}. The above holonomies correspond to the decomposition of the 4-dimensional torus $T^4$ in $T^4=T^2\times T^2$. Therefore, by proceeding in the same way as in the case $D=3$, we see that these holonomies lead to a $(2k_1\times 2k_2)$-fold degeneracy in each layer. For $N$ layers, we get
\begin{equation}
(2k_1 \times 2 k_2)^{N}.
\label{eq65}
\end{equation}
Finally, considering the $2^4$ sub-dimensional manifolds, it follows that the total ground state degeneracy is
\begin{equation}
\text{GSD}= (2k_1 \times 2 k_2)^{2^4 N}=\left[2^2\;\text{Pf}(K)\right]^{2^4 N},
\label{eq66}
\end{equation}
which is expressed in a basis-independent way in terms of the Pfaffian of the original matrix $K$. 

The generalization to the odd $D$-dimensional case is immediate. We decompose the space in a $(D-1)$-dimensional sub-manifold corresponding to one of the  $2^{D-1}$ sub-spaces where charge is conserved, and a perpendicular dimension which is then discretized. Next,  we make the transformation (\ref{eq2a}) to bring the matrix $K$ to the block-diagonal form
\begin{equation}
Q\,K\,Q^{T}=\text{Diag}\left\{
\left(
\begin{array}{cc}
0& k_1\\
-k_1& 0\\
\end{array}
\right),
\left(
\begin{array}{cc}
0& k_2\\
-k_2& 0\\
\end{array}
\right),\ldots,
\left(
\begin{array}{cc}
0& k_{\frac{D-1}{2}}\\
-k_{\frac{D-1}{2}}& 0\\
\end{array}
\right)
\right\},
\label{eq67}
\end{equation}
where all $k$'s are real and positive. In this basis, the fields $A_a$ decouple pairwise, which is equivalent to decomposing the $(D-1)$-dimensional torus as
\begin{equation}
T^{D-1}=\underbrace{T^2\times T^2\times \cdots \times T^2}_{\frac{D-1}{2}}.
\label{eq68}
\end{equation}
The corresponding degeneracy is 
\begin{equation}
2k_1\times 2 k_2\times \cdots \times 2 k_{\frac{D-1}{2}}= 2^{\frac{D-1}{2}}\;\text{Pf}(K)\,.
\label{eq69}
\end{equation}
Taking into account the $N$ layers, we have
\begin{equation}
\left[2^{\frac{D-1}{2}}\;\text{Pf}(K)\right]^{N}\,.
\label{eq70}
\end{equation}
Finally, considering all the $2^{D-1}$ sub-dimensional manifolds, we obtain the total ground state degeneracy
\begin{equation}
\text{GSD}=\left[2^{\frac{D-1}{2}}\;\text{Pf}(K)\right]^{2^{D-1}N}.
\label{eq71}
\end{equation}

For the case of Clifford fractons, where $k_1=k_2=\cdots=k_{\frac{D-1}{2}}=1$ or, equivalently, $\text{Pf}(K)=1$, the ground state degeneracy reduces to 
\begin{equation}
\text{GSD}=2^{(D-1)2^{D-3}L},
\label{eq72}
\end{equation}
where we have again used that $N= L/2$. This is precisely the result shown in the end of Sec. \ref{microscopic model} obtained directly from the lattice model.


\section{Final Remarks}\label{conclusions}

In this work we constructed fracton models on the lattice and
identified their continuum description in terms of Chern-Simons-like
theories. The construction is generic in that it applies to any system
whose microscopic Hamiltonian is a sum of commuting projectors built from tensor products of spin-1/2
operators. Instead of working directly with tensor products of Pauli
operators that represent the local variables, we utilize the Dirac
representation of Clifford algebras. This representation makes a
connection between the lattice model and the field theory simple. Our formalism can, in principle, be used to analyze other lattice models, such as those that exhibit subsystem symmetry protected topological (SSPT) phases~\cite{Chen2019a, You2018} or type II fracton phases~\cite{Haah2011}. Applying this formalism to these problems is a natural direction for future work.

In the field theory, the algebraic structure of the Dirac matrices is
encoded in an anti-symmetric matrix $K$. The details about an specific
lattice model enter via this matrix $K$ (whose dimension depends on
the size of the representation), the charge vectors $T$ (that specify
the operators that are placed on the sites), as well as the lattice
vector positions of the sites themselves. Given these data, one can
follow the prescription here presented and derive an effective field
theory for any type of Clifford-like fracton, such as the 3D Chamon
(with a $2\times 2$ Dirac representation) or the 3D Haah (with a
$4\times 4$ Dirac representation) codes. As a concrete example, we
built fracton theories in odd $D$ spatial dimensional spaces. We discussed the
properties of the resulting Chern-Simons-like theory, such as their
currents, which are conserved in sub-manifolds, and the topological
degeneracy of the ground states, which formally depends on the
Pfaffian of the matrix $K$ and, as usual in fracton systems, on the linear size of the system. 

Properties such as the mutual statistics of the quasiparticles where not explored in the present work. The restricted mobility of fractons makes it unnatural to speak of standard braiding. However, the authors in \cite{PaiHermele2019} were able to develop a theory of fusion and statistical processes that incorporates the mobility restrictions common in fracton models. An interesting question for future exploration is how our formalism could incorporate their notion of statistics.

For readers familiar with the $K$-matrices and charge vectors $T$
appearing in the description of Abelian fractional quantum Hall
states \cite{Wen1995}, as well as their quantum wire constructions \cite{Kane2002, Teo2014, Neupert2014, Fontana2019}, it is tempting to
expect that the description here presented -- for ``integer'' fractons
given our $K$ and $T$'s -- could possibly lend itself to the analysis
of fractional fractons. This is an intriguing possibility that merits
further investigation, but keeping the following points in mind.

The approach of this paper resembles quantum wire constructions of
topological phases, but instead of wires we deploy $(0+1)$-dimensional
degrees of freedom, i.e., ours is a ``quantum dot'' construction. Like in
the wire constructions, we identify families of commuting operators
that can be simultaneously pinned and gap the system. In the wire
systems, fractionalization already takes place in the
$(1+1)$-dimensional building blocks, and it is carried over to higher dimensions by coupling the
wires, notably using only integer charge transfer operators. However,
there is no fractionalization in the quantum dots of the construction
of this paper. Of course, one may generalize the construction presented here to
start with wires instead of dots, in which case fractionalization may appear
more easily.

\paragraph*{Added note}: It has been brought to our attention that the word "fracton" has been used in physics in other contexts before. An early use was in \cite{Khlopov1981} in reference to fractional charges in quantum chromodynamics. In our construction, we adopt the modern meaning of the word as stressed in the main text. 

\section*{Acknowledgements}
This work is supported by the Brazilian agency Coordenação de
Aperfeiçoamento de Pessoal de Nível Superior (CAPES) under grant
number 88881.361635/2019-01 (W.~F.), the CNPq grant number
311149/2017-0 (P.~G.), and the DOE Grant
No. DE-FG02-06ER46316 (C~.C). W.~F. acknowledges support by the
Condensed Matter Theory Visitors program at Boston University.

\begin{appendix}
	\section{Euclidean Dirac matrix representations of Clifford algebras}
	\label{AA}
	
	We construct fracton models in odd $D=2n+1$ dimensions using
	representations of the Clifford algebra. Specifically, we use the
	Euclidean Dirac matrices. Below we construct these representations and
	show properties that these matrices satisfy. These properties are
	used, for example, to argue that there is no operator that can move
	defects in the corresponding fracton models.
	
	Let us work with matrices defined as the tensor products of Pauli matrices:
	\begin{align}
	\label{eq:prod-Paulis}
	\gamma^{(n)}_{\mu_1\,\mu_2\,\dots\mu_n}
	\equiv
	\sigma_{\mu_1} \otimes \sigma_{\mu_2} \otimes \dots\otimes  \sigma_{\mu_n}
	\;,
	\end{align}
	with $\mu_i=0,1,2,3$ and $\sigma_0\equiv\openone$. We shall obtain a set of $2n+1$ mutually anticommuting matrices for
	any $n$. We construct this set inductively.
	
	For $n=1$, the set contains the matrices $\gamma^{(1)}_1=\sigma_1$,
	$\gamma^{(1)}_2=\sigma_2$, and $\gamma^{(1)}_3=\sigma_3$. Equivalently, we can label these
	matrices as $\gamma^{(1)}_I$, with indices $I\in S^{(1)}=\{1,2,3\}$.
	
	For $n=2$, we first construct the following 3 matrices using the
	$\gamma^{(1)}_I$, $I\in S^{(1)}$: $\gamma^{(2)}_{I3} = \gamma^{(1)}_{I}\otimes
	\sigma_3$. Second, we take the following two matrices: $\gamma^{(2)}_{01}$
	and $\gamma^{(2)}_{02}$. Therefore the five matrices $\gamma^{(2)}_I$, with
	indices $I\in S^{(2)}=\{13,23,33,01,02\}$, are all anticommuting.
	
	We proceed by induction. Suppose that we have $2n-1$ anticommuting
	matrices $\gamma^{(n-1)}_{i}$, $i\in S^{(n-1)}$. First, using the $2n-1$
	matrices $\gamma^{(n-1)}_i$, $i\in S^{(n-1)}$, build the matrices
	$$\gamma^{(n)}_{i3}
	= \gamma^{(n-1)}_{i}\otimes \sigma_3\;.$$
	Second, take the two matrices
	$$
	\gamma^{(n)}_{0\dots 0\,1}
	\quad
	\text{and}
	\quad
	\gamma^{(n)}_{0\dots 0\,2}\;.
	$$
	The $2n-1+2=2n+1$ matrices $\gamma^{(n)}_{I}$, with $I\in S^{(n)} =
	\{i3 \;|\; i\in S^{(n-1)}\}\cup \{0\dots 0\,1, \;0\dots 0\,2\}$ are all
	anticommuting.
	
	These $2n+1$ matrices multiply to the identity up to a prefactor:
	\begin{align}
	\prod_{I\in S^{(n)}} \;\gamma^{(n)}_{I} = \pm i^n\;\;\gamma^{(n)}_{0\dots 0\,0}
	\;,
	\end{align}
	where the $\pm$ simply depends on the order that the matrices are
	multiplied (the choice of order of the indices $I\in S^{(n)}$). This
	relation can also be proved by induction. Notice that it holds for
	$n=1$. If it holds for $n-1$, then it follows that
	\begin{align}
	\prod_{I\in S^{(n)}} \;\gamma^{(n)}_{I}
	&=
	\left(\prod_{i\in S^{(n-1)}} \;\gamma^{(n)}_{i3}\right)\;\gamma^{(n)}_{0\dots 0\,1}\;\gamma^{(n)}_{0\dots 0\,2}
	\nonumber\\
	&=
	\left(\pm i^{n-1}\;\gamma^{(n)}_{0\dots 0\,3}\right)\;\gamma^{(n)}_{0\dots 0\,1}\;\gamma^{(n)}_{0\dots 0\,2}
	\nonumber\\
	&=
	\mp i^{n} \;\gamma^{(n)}_{0\dots 0\,0}
	\;.
	\end{align}
	This property means that the last, or $(2n+1)$th, $\gamma$-matrix can
	be obtained from the product of all the other $2n$ matrices. It also
	follows that any matrix that is a tensor product of Pauli matrices can
	be written as products of these $2n$ $\gamma$-matrix. (Notice that
	there are $4^n$ possible tensor products of Pauli matrices, and
	$2^{2n}=4^n$ choices of whether a $\gamma$-matrix enters or not the
	product of $\gamma$'s.)
	
	The construction above yields a set of $2n+1$ matrices
	$\gamma^{(n)}_{I}$ satisfying
	\begin{align}
	\{\gamma^{(n)}_{I},\gamma^{(n)}_{J}\}=2\,\delta_{IJ}\;.
	\end{align}
	The set of indices $I\in S^{(n)}$ can be interchanged to $I=1,\dots,2n+1$, which
	is the notation we use in the main text for the Euclidean Dirac
	matrices.

	
	\subsection{Properties of the Euclidean Dirac matrices}
	\label{sec:properties}
	
	Let us now show three useful properties of the $2n+1$ matrices $\gamma^{(n)}_{I}$ with $I\in S^{(n)}$.
	
	\begin{enumerate}
		
		
		\item \textbf{The identity is the only tensor product of Pauli matrices that
			commutes with all the Dirac matrices}. In other words, only the
		matrix $\gamma^{(n)}_{J}$, $J= 00\dots0$, can commute the $2n+1$
		matrices $\gamma^{(n)}_{I}$ with $I\in S^{(n)}$.
		
		To show this property, suppose that there is a matrix $\gamma^{(n)}_{J}$
		that commutes with all the $2n+1$ matrices. This $J$ must be of the
		form $J=j0$ for $\gamma^{(n)}_{J}$ to commute with both $\gamma^{(n)}_{0\dots
			0\,1}$ and $\gamma^{(n)}_{0\dots 0\,2}$. Therefore,
		$$
		[\gamma^{(n)}_{J},\, \gamma^{(n)}_{I}]=0,
		\quad
		\forall I\in S^{(n)}
		\quad
		\Leftrightarrow
		\quad
		[\gamma^{(n-1)}_{j},\, \gamma^{(n-1)}_{i}]=0,
		\quad
		\forall i\in S^{(n-1)}
		\;.
		$$
		We can use this recursion all the way to $n=1$, where only $\gamma^{(1)}_0$
		commutes with the $\gamma^{(1)}_i, i\in S^{(1)}$, and conclude that $J$
		must be $J= 00\dots0$, i.e., all the entries must be 0.
		
		
		
		\item \textbf{The set of matrices $\gamma^{(n)}_{I}$ with $I\in S^{(n)}$ is
			maximal, i.e., no other matrix can be added to the set that
			anticommutes with those already in}.
		The statement is true for $n=1$: the matrices $\gamma^{(1)}_{I}$ with $I\in
		S^{(1)}$ are the three Pauli matrices, leaving no other option to
		include that would anticommute with these three.
		
		Now suppose that the statement is true up to $n-1$; let us analyze the
		consequences for when we consider $n$.
		
		Suppose by contradiction that there exists a $J\notin S^{(n)}$ such
		that $\gamma^{(n)}_{J}$ anticommutes with all the $\gamma^{(n)}_{I}$ with $I\in
		S^{(n)}$. Let us break the problem in four cases, and show
		impossibility in all cases.
		\begin{itemize}
			
			\item $J=j0$
			
			This is the simplest case: $\gamma^{(1)}_{j0}$ commutes with both
			$\gamma^{(n)}_{0\dots 0\,1}$ and $\gamma^{(n)}_{0\dots 0\,2}$, so $J=j0$
			cannot be added to the set.
			
			\item $J=j1$ (the case $J=j2$ is analogous)
			
			This case is also simple: $\gamma^{(1)}_{j1}$ commutes with
			$\gamma^{(n)}_{0\dots 0\,1}$, so $J=j1$ cannot be added to the set.
			
			\item $J=j3$
			
			$\gamma^{(1)}_{j3}$ anticommutes with $\gamma^{(n)}_{0\dots 0\,1}$ and
			$\gamma^{(n)}_{0\dots 0\,2}$, so we should only consider the
			anticommutation with the other $2n-1$ matrices $\gamma^{(n)}_{i3}$, for
			$i\in S^{(n-1)}$. But because $S^{(n-1)}$ is maximal, there is no
			new $j\notin S^{(n-1)}$ to add.
			
		\end{itemize}
		
		We thus conclude that the set of $2n+1$ matrices $\gamma^{(n)}_{I}$ with
		$I\in S^{(n)}$ is maximal.

		
		
		\item \textbf{There is no matrix $\gamma^{(n)}_{J}$ that
			commutes with $2n$ of the matrices $\gamma^{(n)}_{I}$ with $I\in
			S^{(n)}$. Therefore, defects cannot be created in only one direction}. Phrasing it differently, this states that there is no line defects on the model. This result will allow us to argue that we can construct a
		fracton model.
		
		The statement is true for $n=1$: there is no matrix $\gamma^{(1)}_{J}$ that
		commutes with two of the matrices $\gamma^{(1)}_{I}$ with $I\in S^{(1)}$,
		because no one Pauli matrix commutes with two Pauli matrices.
		
		Now suppose that the statement is true up to $n-1$; let us analyze the
		consequences for when we consider $n$.
		
		Let us break the problem in four cases:
		\begin{itemize}
			
			\item $J=j0$
			
			In this case, the commutation with $\gamma^{(n)}_{0\dots 0\,1}$ and
			$\gamma^{(n)}_{0\dots 0\,2}$ comes for free. Therefore we reduce the
			problem to finding $\gamma^{(n-1)}_{j}$ that commutes with $2(n-1)$
			matrices $\gamma^{(n-1)}_{i}$ with $i\in S^{(n-1)}$. Since there is no
			solution for this problem (the statement is true for the case with
			$n-1$), then there is no solution for the case with $n$ either.
			
			\item $J=j3$
			
			This is the simplest case; $\gamma^{(1)}_{j3}$ anticommutes with
			$\gamma^{(n)}_{0\dots 0\,1}$ and $\gamma^{(n)}_{0\dots 0\,2}$, so it is
			impossible that there are $2n$ other matrices that commute with
			$\gamma^{(n)}_{J}$ among the $\gamma^{(n)}_{I}$ with $I\in S^{(n)}$, since
			there are at most $2n+1-2=2n-1<2n$.
			
			\item $J=j1$ (the case $J=j2$ is analogous)
			
			$\gamma^{(1)}_{j1}$ commutes with $\gamma^{(n)}_{0\dots 0\,1}$ and
			anticommutes with $\gamma^{(n)}_{0\dots 0\,2}$. So we need to find $2n-1$
			additional matrices that commute with $\gamma^{(n)}_{j1}$ among the
			$\gamma^{(n)}_{i3}$ with $i\in S^{(n-1)}$. This is equivalent to finding
			$2n-1$ matrices that anticommute with $\gamma^{(n-1)}_{j}$ among the
			$\gamma^{(n-1)}_{i}$ with $i\in S^{(n-1)}$. This is impossible since the
			set $S^{(n-1)}$ is maximal (see above).
			
		\end{itemize}
		
		
	\end{enumerate}


	\section{Fracton models build from the Clifford algebra representations}
	\label{AB}
	
	We can construct a fracton model in $D$ dimensions if $D$ is odd. In
	this case we take $n=(D-1)/2$ and we can use the matrices
	$\gamma^{(n)}_{I}$, $I\in S^{(n)}$ in the construction. Here we shall
	label these $2n+1$ matrices simply $\gamma^I, I=1,\dots,2n+1$, as we
	did in the main text.
	
	The construction can be made in the $D$-dimensional hypercube, with
	orthogonal basis vectors $\hat a_I, I=1,\dots,D$, as presented in the
	main text. We place the degrees of freedom on the even sublattice
	$\Lambda_e$. The dimension of the local Hilbert space at each site is
	$2^n$, or equivalently, that associated with the $n$ spins or gradings
	of Pauli operators used to construct the $\gamma$-matrix
	representations. At these even sublattice sites we place operators
	$\Gamma^{(I,\alpha)}, \alpha=1,\dots,n$ built as products of the
	$\gamma$-matrices (in turn built from tensor products of Pauli
	matrices).
	
	The first set of operators, with $\alpha=1$, is the set of Dirac
	matrices $\gamma^I, I=1,\dots,2n+1$, or explicitly
	\begin{align}
	\label{eq:1st-set}
	\Gamma^{(I,1)} = \gamma^I
	\;.
	\end{align}
	The other sets are needed to gap the model.
	
	We define $\mathcal{O}^{(\alpha)}_{\vec x}$ operators centered at sites $\vec
	x$ on the odd sublattice $\Lambda_o$,
	\begin{equation}
	\label{eq1A}
	\mathcal{O}^{(\alpha)}_{\vec{x}}
	\equiv
	\prod^D_{I=1} \;\;
	\Gamma^{(I,\alpha)}_{\vec{x}-\hat{a}_I}
	\;\;
	\Gamma^{(I,\alpha)}_{\vec{x}+\hat{a}_I}
	\;,
	\quad
	\alpha=1,\dots,\frac{(D-1)}{2}
	\;,
	\end{equation}
	and using these the Hamiltonian 
	\begin{equation}
	\label{eq1bA}
	H=-\sum_{\alpha=1}
	\;
	\left(g_\alpha \;\sum_{\vec{x}}\mathcal{O}^{(\alpha)}_{\vec{x}} \right)
	\;.
	\end{equation}
	We can choose the operators $\Gamma^{(I,\alpha)}$ such that
	\begin{equation}
	\label{eq1c-appendix}
	\left[
	\mathcal{O}^{(\alpha)}_{\vec{x}}
	\;,\;
	\mathcal{O}^{(\beta)}_{\vec{x}'}
	\right]
	=
	0\;,
	\quad
	\forall \alpha,\beta\;, \forall {\vec{x}},{\vec{x}'}
	\;.
	\end{equation}
	As stated in the main text, in this case \textit{i}) the Hamiltonian is a sum
	of commuting projectors, and \textit{ii}) there are as many commuting
	projectors as the number of degrees of freedom in the problem (up to
	constraints tied to the topological degeneracy).
	
	Let us first focus on the operators $\mathcal{O}^{(1)}$ for simplicity.
	These are defined as
	\begin{align}
	\mathcal{O}^{(1)}_{\vec x}=
	\prod_{I=1}^D \;\gamma^{I}_{\vec x-\hat a_I}\;\gamma^{I}_{\vec x+\hat a_I}
	\;.
	\end{align}
	The operators $\gamma^{I}_{\vec x}$ satisfy the following commutation
	relations:
	\begin{align}
	\{\gamma^{I}_{\vec x}\;,\;\gamma^{J}_{\vec x'}\}=2\;\delta_{IJ}
	\;, \;\text{if}\;\; \vec x = \vec x'\;,
	\qquad\text{and}\qquad
	[\gamma^{I}_{\vec x}\;,\;\gamma^{J}_{\vec x'}]=0
	\;, \;\text{if}\;\; \vec x \ne \vec x'\; 
	\;.
	\end{align}
	(We remark that these models are bosonic, and not fermionic; the Dirac
	matrices represent the tensor product of local Pauli matrices, that in
	turn represent spin degrees of freedom on the lattice.) Given these
	commutation relations, it follows that all distinct
	$\mathcal{O}^{(1)}_{\vec x}$ and $\mathcal{O}^{(1)}_{\vec x'}$ that
	share common sites commute: 1) they either share a single site along
	the line that connects them, in which case the same operator (same
	$I$), or 2) they share two sites with different components $I$ and $J$
	entering in each of $\mathcal{O}^{(1)}_{\vec x}$ and
	$\mathcal{O}^{(1)}_{\vec x'}$, and hence there is a factor of $-1$
	from the anti-commutation relation of each common site, and hence in
	total a factor $(-1)^2$, leading to the commutation of the two
	operators.
	
	The operators $\mathcal{O}^{(1)}_{\vec x}$ square to unity, and thus
	have eigenvalues $\pm 1$. The ground state has all eigenvalues +1 for
	all operators. Excitations correspond to eigenvalues -1. Because we
	used all the $2n+1$ Dirac matrices in constructing
	$\mathcal{O}^{(1)}_{\vec x}$, and as demonstrated above in
	Sec.~\ref{sec:properties}, there is {\it no} operator that
	anti-commutes with one and only one of the $\gamma^I$. Therefore, it
	is not possible to construct a local operator whose sole effect is to
	create a pair of defects, or move a single defect. Defects are only
	created in at least quadruplets in any dimension $D=2n+1$, much as in
	the $D=3$ model in Ref.~\cite{Chamon2005}. This property that defects
	cannot be created in pairs, but only in at least quadruplets,
	underscores the fracton nature of these odd $D$ models.
	
	Let us now discuss the other operators
	$\mathcal{O}^{(\alpha)}_{\vec{x}}$, $\alpha=2,\dots,n$. The argument
	for the commutativity follows a similar line. When two operators
	$\mathcal{O}^{(\alpha)}_{\vec x}$ and $\mathcal{O}^{(\beta)}_{\vec
		x'}$ share sites, there are two cases to consider.
	
	The case when they share one site: the neighboring $\mathcal{O}$'s,
	defined at sites $\vec{x}$ and $\vec{x}+2\hat{a}_I$ of $\Lambda_o$,
	share the $\Lambda_e$ site at $\vec{x}+\hat{a}_I$, and they commute if
	\begin{align}
	\left[
	\Gamma^{(I,\alpha)}
	\;,\;
	\Gamma^{(I,\beta)}
	\right]
	=0
	\;.
	\end{align}
	
	The case when they share two sites: the neighboring
	$\mathcal{O}^{(\alpha)}$'s, defined at sites $\vec{x}$ and
	$\vec{x}+\hat{a}_I+\hat{a}_J$ of $\Lambda_o$, share the two
	$\Lambda_e$ sites at $\vec{x}+\hat{a}_I$ and $\vec{x}+\hat{a}_J$. The
	operators on those sites either commute or anti-commute, which can be
	cast as
	\begin{align}
	\label{eq:eta-def}
	\Gamma^{(I,\alpha)}\;\;\Gamma^{(J,\beta)}
	=
	(-1)^{\eta^{(\alpha\beta)}_{IJ}}\;\;
	\Gamma^{(J,\beta)}\;\;\Gamma^{(I,\alpha)}
	\;,
	\end{align}
	with $\eta^{(\alpha\beta)}_{IJ}=0$ or $1$, and the desired commutation
	relations Eq.~(\ref{eq1c-appendix}) are guaranteed if
	\begin{align}
	\label{eq:eta-equality}
	\eta^{(\alpha\beta)}_{IJ}
	=
	\eta^{(\alpha\beta)}_{JI}
	\;.
	\end{align}
	This condition, or equivalently that $\eta^{(\alpha\beta)}_{IJ}+
	\eta^{(\alpha\beta)}_{JI}=0\mod 2$, is the counterpart to
	$C^{(\alpha\beta)}_{IJ}=0$ of Eq.~(\ref{eq13a}) in the main text. The
	components $T^{(I,\alpha)}$ of the $T$-vectors, when computed mod 2,
	simply encode which of the $\gamma^I$ matrices enter in the product
	defining the operator $\Gamma^{(I,\alpha)}$. Because of this relation,
	we shall show how to construct the $\Gamma^{(I,\alpha)}$'s by showing
	how to ensure $C^{(\alpha\beta)}_{IJ}=0$, which we can solve more
	easily using integer instead of binary vectors.
	
	Explicitly, we construct the operators $\Gamma^{(I,\alpha)}$ using
	$2n$-dimensional $T$-vectors, $T^{(I,\alpha)}_a, a=1,\dots,2n$, as
	follows:
	\begin{align}
	\Gamma^{(I,\alpha)}
	=
	\left(\gamma^1\right)^{T^{(I,\alpha)}_1}\;
	\left(\gamma^2\right)^{T^{(I,\alpha)}_2}\;
	\dots\;
	\left(\gamma^{2n}\right)^{T^{(I,\alpha)}_{2n}}\;
	\;.
	\end{align}
	Notice that since $(\gamma^I)^2=1$, only the values of the $T$-vectors
	mod 2 matter. The particular case of the first set, see
	Eq.~(\ref{eq:1st-set}), corresponds to the vector
	\begin{align}
	\label{eq:1st-set}
	T^{(I,1)}_a=t^I_a=\delta^I_a,\;\; I=1,\dots,2n\;,
	\quad {\rm and}\quad
	T^{(2n+1,1)}_a=-\sum_{I=1}^{2n}\;t^{I}_a
	\;.
	\end{align}
	(The $t^I$ are the basis vectors.)
	
	The commutation relations between the $\gamma$-matrices can be encoded
	in an integer-valued anti-symmetric $K$-matrix via
	\begin{align}
	\gamma^{I}\;\;\gamma^{J}
	=
	e^{i\pi\;t^{(I)}_a\;K_{ab}\;t^{(J)}_b}\;
	\gamma^{J}\;\;\gamma^{I}
	\;,
	\end{align}
	where repeated index summation over the $a$ and $b$ are used. The
	correct commutation relations follow from requiring that
	\begin{align}
	t^{(I)}_a\;K_{ab}\;\,t^{(J)}_b
	=
	\begin{cases}
	0 \mod 2, \quad I=J
	\\
	1 \mod 2, \quad I\ne J
	\end{cases}
	\;.
	\end{align}
	
	It follows that the commutation relations
	\begin{align}
	\Gamma^{(I,\alpha)}\;\;\Gamma^{(J,\beta)}
	=
	e^{i\pi\;T^{(I,\alpha)}_a\,K_{ab}\;\,T^{(J,\beta)}_b}\;\;\;
	\Gamma^{(J,\beta)}\;\;\Gamma^{(I,\alpha)}
	\;,
	\end{align}
	or equivalently, using Eq.~(\ref{eq:eta-def}),
	\begin{align}
	\eta^{(\alpha\beta)}_{IJ}
	=
	T^{(I,\alpha)}_a\,K_{ab}\;\;T^{(J,\beta)}_b
	\mod 2
	\;.
	\end{align}
	Then, condition Eq.~(\ref{eq:eta-equality}) is equivalent to
	\begin{align}
	C^{(\alpha\beta)}_{IJ} = 0 \mod 2
	\;,
	\end{align}
	where
	\begin{align}
	C^{(\alpha\beta)}_{IJ} \equiv
	T^{(I,\alpha)}_a\,K_{ab}\;\;T^{(J,\beta)}_b
	+
	T^{(J,\alpha)}_a\,K_{ab}\;\;T^{(I,\beta)}_b
	\;.
	\end{align}
	While we just need $C^{(\alpha\beta)}_{IJ}$ to vanish mod 2, we can
	simply demand that it vanishes, and still solve the problem as we
	show below.
	
	Let us now construct vectors $T^{(I,\,\alpha)}_a$ that satisfy
	$C^{(\alpha\beta)}_{IJ}=0$. We already have the first set of
	$T$-vectors from Eq.~(\ref{eq:1st-set}). Now build the other sets of
	$T$-vectors via a family of linear transformations $L^{(\alpha)}$:
	\begin{align}
	T^{(I,\alpha)}_a=\sum_M L^{(\alpha)}_{IM}\;T^{(M,1)}_a
	=
	L^{(\alpha)}_{Ia}\;,\; I,M=1,\dots,2n\;,
	\quad {\rm and}\quad
	T^{(2n+1,\alpha)}_a=-\sum_{I=1}^{2n}\;T^{(I,\alpha)}_a
	\;.
	\end{align}
	It follows, for $I,J = 1,\dots,2n$, that
	\begin{align}
	C^{(\alpha\beta)}_{IJ}
	&=
	T^{(I,\,\alpha)}_a\; K_{ab}\; T^{ (J,\,\beta)}_{b}
	+
	T^{(J,\,\alpha)}_a\; K_{ab}\; T^{ (I,\,\beta)}_{b}
	\nonumber\\
	&=
	L^{(\alpha)}_{Ia}\; K_{ab}\; L^{(\beta)}_{Jb}
	+
	L^{(\alpha)}_{Ja}\; K_{ab}\; L^{(\beta)}_{Ib}
	\nonumber\\
	&=
	({L^{(\alpha)}}^{}\; K\; {L^{(\beta)}}^\top)_{IJ}
	+
	({L^{(\beta)}}^{}\; K^\top\; {L^{(\alpha)}}^\top)_{IJ}
	\;,
	\end{align}
	or equivalently, that
	\begin{align}
	C^{(\alpha\beta)}
	&=
	{L^{(\alpha)}}^{}\; K\; {L^{(\beta)}}^\top
	+
	{L^{(\beta)}}^{}\; K^\top\; {L^{(\alpha)}}^\top
	\nonumber\\
	&=
	{L^{(\alpha)}}^{}\; K\; {L^{(\beta)}}^\top
	+
	({L^{(\alpha)}}^{}\; K\; {L^{(\beta)}}^\top)^\top
	\;.
	\label{eq:condition}
	\end{align}
	Hence the condition that the commutation relations
	$C^{(\alpha\beta)}$ vanish require that the sets of
	$(\alpha,\beta)$-indexed matrices $({L^{(\alpha)}}^{}\; K\;
	{L^{(\beta)}}^\top)$ be anti-symmetric (in the indices $I$ and $J$).
	Let then
	\begin{align}
	{L^{(\alpha)}}^{}\; K\; {L^{(\beta)}}^\top
	=
	A^{(\alpha\beta)}
	\;,
	\end{align}
	where the $A^{(\alpha\beta)}$ are anti-symmetric matrices for any of the
	$\alpha,\beta$ pairs. For given choices of matrices $A^{(\alpha\beta)}$,
	we can solve sequentially for
	\begin{align}
	{L^{(\alpha)}}^{}
	=
	A^{(\alpha\beta)}\;({L^{(\beta)}}^\top)^{-1}\;K^{-1}
	\;,
	\label{eq:Ls}
	\end{align}
	i.e., start with $\beta=1$ and $L^{(1)}=\openone$, obtain $L^{(2)}$
	for some arbitrary choice of $A^{(21)}$, then for some choice
	$A^{(31)}$ obtain $L^{(3)}$, and so on. In other words, we can
	determine the ${L^{(\alpha)}}^{}$ from using $\beta=1$ and
	$L^{(1)}=\openone$ in Eq.~(\ref{eq:Ls}):
	\begin{align}
	{L^{(\alpha)}}^{}
	=
	A^{(\alpha\,1)}\;K^{-1}
	\;,
	\label{eq:Ls-sol}
	\end{align}
	for anti-symmetric choices of $A^{(\alpha\,1)}$. Notice that the
	${L^{(\alpha)}}^{}$ cannot be equal, otherwise two sets of
	$T^{(I,\alpha)}$'s would be identical. The number of solutions (number
	of $\alpha's$) depend on the dimension $D-1$ of the matrices, for
	example the matrix $K$. Notice that if $K$ is $2\times 2$, any
	anti-symmetric matrix is proportional to $i\sigma_2$, and therefore it
	follows from Eq.~(\ref{eq:Ls-sol}) that one cannot get a non-trivial
	solution other than $L^{(1)}\propto\openone$.
	
	There are compatibility conditions for the matrices, because one can
	reach, for example, $L^{(3)}$ from $L^{(1)}$ or $L^{(2)}$. For example,
	\begin{align}
	{L^{(\alpha)}}^{}\; K\; {L^{(\beta)}}^\top
	&=
	A^{(\alpha\,1)}\;K^{-1}\;K\;{K^{-1}}^\top {A^{(\beta\,1)}}^\top
	\nonumber\\
	&=
	A^{(\alpha\,1)}\;K^{-1}\;{A^{(\beta\,1)}}
	\end{align}
	or equivalently
	\begin{align}
	\label{eq:compatibility}
	A^{(\alpha\beta)}
	&=
	A^{(\alpha\,1)}\;K^{-1}\;{A^{(\beta\,1)}}
	\;.
	\end{align}

	\subsection{Example of $D=5$}
	\renewcommand*{\arraystretch}{1.5}
	
	Consider the following $4\times 4$ $K$-matrix:
	\begin{align}
	K_4
	=
	\begin{bmatrix}
	\;0&+1&+1&+1\\
	-1&\;0&+1&+1\\
	-1&-1&\;0&+1\\
	-1&-1&-1&\;0
	\end{bmatrix}
	\;,
	\end{align}
	with inverse
	\begin{align}
	K_4^{-1}
	=
	\begin{bmatrix}
	\;0&-1&+1&-1\\
	+1&\;0&-1&+1\\
	-1&+1&\;0&-1\\
	+1&-1&+1&\;0
	\end{bmatrix}
	\;.
	\end{align}
	The choice $A^{(1\,1)}=K_4$ yields $L^{(1)}=\openone$, as it should
	be. Choose the anti-symmetric matrix $A^{(2\,1)}=K_2\otimes \openone_2$,
	where $K_2=\big(\begin{smallmatrix} 0 & +1\\ -1 &
	0 \end{smallmatrix}\big)$, or explicitly,
	\begin{align}
	A^{(2\,1)}
	=
	\begin{bmatrix}
	\;0&0&+1&0\\
	0&\;0&0&+1\\
	-1&0&\;0&0\\
	0&-1&0&\;0
	\end{bmatrix}
	\;,
	\end{align}
	from which we obtain
	\begin{align}
	L^{(2)} =
	\begin{bmatrix}
	\;-1&+1&\;0&-1\\
	+1&-1&+1&\;0\\
	\;0&+1&-1&+1\\
	-1&\;0&+1&-1
	\end{bmatrix}
	\;.
	\end{align}
	From the $L$ matrix we obtain the vectors
	\begin{align}
	\label{B30}
	{ T}^{(1,2)} &= (-1,+1,\;0,-1)
	\nonumber\\
	{ T}^{(2,2)} &= (+1,-1,+1,\;0)
	\nonumber\\
	{ T}^{(3,2)} &= (\;0,+1,-1,+1)
	\nonumber\\
	{ T}^{(4,2)} &= (-1,\;0,+1,-1)
	\nonumber\\
	{ T}^{(5,2)} &= (+1,-1,-1,+1)        
	\;.
	\end{align}
	The corresponding operators $\Gamma^{(I,2)}$ are:
	\begin{align}
	\Gamma^{(1,2)} &= \gamma^1\;\gamma^2\;\gamma^4 \;\;\sim\;\; \gamma^3\;\gamma^5
	\nonumber\\
	\Gamma^{(2,2)} &= \gamma^1\;\gamma^2\;\gamma^3 \;\;\sim\;\; \gamma^4\;\gamma^5
	\nonumber\\
	\Gamma^{(3,2)} &= \gamma^2\;\gamma^3\;\gamma^4 \;\;\sim\;\; \gamma^1\;\gamma^5
	\nonumber\\
	\Gamma^{(4,2)} &= \gamma^1\;\gamma^3\;\gamma^4 \;\;\sim\;\; \gamma^2\;\gamma^5
	\nonumber\\
	\Gamma^{(5,2)} &= \gamma^1\;\gamma^2\;\gamma^3\;\gamma^4 \;\;\sim\;\;\gamma^5
	\;.
	\end{align}
	
	One can summarize the operators $\Gamma^{(I,\alpha)}$ in the following
	table, as we did in the main text:
	\begin{align}
	\begin{matrix}
	\Gamma^{(I,\alpha)} &\vline& I =& 1 & 2 & 3 & 4 & 5 &
	\\
	\hline
	\alpha=1&\vline& & \gamma^1 & \gamma^2 & \gamma^3 & \gamma^4 & \gamma^5 &
	\\
	\alpha=2&\vline& &
	\gamma^3 \gamma^5\; & \gamma^4 \gamma^5\; & \gamma^1 \gamma^5 \;&
	\gamma^2 \gamma^5 \; & \gamma^5 &
	\end{matrix}
	\end{align}

	\subsection{Construction for general $D=2n+1$}
	
	Define the following $n\times n$ anti-symmetric matrix:
	\begin{align}
	K_{n} =
	\begin{bmatrix}
	0 & +1 &+1 & \dots & +1\\
	-1 & 0 & +1& \dots & +1\\
	-1 & -1 & 0& \dots & +1\\
	\vdots&\vdots&&\ddots &\vdots& \\
	-1&-1& \dots&-1&0
	\end{bmatrix}_{n\times n}
	\end{align}
	The $2n\times 2n$ $K$-matrix we need for $D=2n+1$ is then simply
	$K_{2n}$.
	
	The following anti-symmetric matrices $A^{(\alpha\,1)}$ 
	\begin{align}
	A^{(1\,1)} & = K_{2n}\,,
	\nonumber\\
	A^{(\alpha\,1)} & = (K_{n})^{2\alpha-3}\otimes \openone_2
	\;,\quad \alpha=2,\dots,n
	\;.
	\end{align}
	commute with both $K_{2n}$ and $K_{2n}^{-1}$; using this property and
	the anti-symmetry of both the $A^{(\alpha\,1)}$ and the $K_{2n}$, one
	can show that the $A^{(\alpha\beta)}$ obtained through
	Eq.~(\ref{eq:compatibility}) are anti-symmetric, as required.
	
	With these $A^{(\alpha\,1)}$, one can proceed to find the
	$L^{(\alpha)}$ matrices and then the vectors $T^{(I,\alpha)}$, and
	finally the operators $\Gamma^{(I,\alpha)}$ with the desired
	commutation relations.
	
	\section{Degeneracy}\label{AC}
	
	In the main text we argued that the topological degeneracy of the
	model is at least $2^{(D-1)\;2^{D-2}}$. This number follows from the
	constraints of multiplying all the $\mathcal{O}^{(\alpha)}$ operators:
	\begin{equation}
	\label{lat constraint a}
	\prod_{\vec{x}\in \Lambda_{o, k}}\mathcal{O}^{(\alpha)}_{\vec{x}} = \openone
	\;,
	\quad k=1,\ldots, 2^{D-1} ,\quad \alpha=1,\dots,(D-1)/2
	\end{equation}
	where $k$ labels the $2^{D-1}$ sub-lattices $\Lambda_{o, k}$. (A unit
	cell of the hypercubic lattice contains $2^D$ sites, half of them are
	on the even and half on the odd sub-lattice -- hence there are
	$2^{D-1}$ distinct sub-lattices of the odd sub-lattice.)
	
	The degeneracy can be greater, and can depend on the system size. Here
	we follow Bravyi, Leemhuis, and Terhal's calculation in their appendix
	A of \cite{Terhal2011}.
	
	It follows from $\sum_{I=1}^{D}\;T^{(I,\alpha)}_a=0$ that the
	$\Gamma^{(I,\alpha)}$ multiply to the identity (up to a factor of
	magnitude 1, that also depends on the order of multiplication). Using
	this property, we arrive at the equivalent of their parity checks:
	\begin{align}
	\bigoplus_{\substack{{J\ne I}\\q =\pm 1}}
	\;t^{(\alpha)}(\vec x+q\,\hat a_J)
	=0
	\;,
	\quad
	\vec x \in \Lambda_e,~~I = 1,2,\dots D
	\;,
	\quad \alpha=1,\dots,\frac{D-1}{2}~
	\;.
	\end{align}
	(The $t$ above conforms to their notation; they are not related to our
	$t$-vectors.) Notice that, for each $\alpha$, the $D$ equations are
	linearly dependent, and that the sum over all of their left hand side
	is identically zero. Suming any pair of these equations yield
	\begin{align}
	t^{(\alpha)}(\vec x-\hat a_I) \oplus
	t^{(\alpha)}(\vec x+\hat a_I) \oplus
	t^{(\alpha)}(\vec x-\hat a_J) \oplus
	t^{(\alpha)}(\vec x+\hat a_J) 
	=0
	\;,
	\quad
	I,J = 1,2,\dots, D
	\;.
	\end{align}
	The solutions of these equations for the case when $L_1=L_2=\dots=L_D$
	in a similar way as in 3-dimensions: first use two lines (with
	$2L/2=L$ sites on a sublattice) and generate the solution on a plane,
	then two planes and generate the solution in the 3rd dimension, and
	after that proceed accordingly, use 2 3-dimensional hyperplanes to
	generate the solutions in 4-dimensions, and so on. The number of
	logical quibts generated in this way is ${\cal C}_D=L/2\times 2\times
	2\times\cdots\times 2$, with $D-1$ 2's, i.e., ${\cal
		C}_D=2^{D-2}\,L$. When we take into account all the
	$\alpha=1,\dots,(D-1)/2$, we have $(D-1)\;2^{D-3}\,L$ logical
	qubits. Therefore, the ground state degeneracy is
	\begin{align}
	\textrm{GSD}
	=
	2^{(D-1)\;2^{D-3}\;L}
	\;.
	\end{align}
	\section{Level quantization from the effective field theory}\label{AF}
	Here we give the details of the calculations in \ref{PropertiesA}. To understand quantization of the matrix $K$ for arbitrary dimensions, it is convenient to consider large gauge transformations depending only on time. In other words, we impose the condition $t\in [0,\tau)$, and place the system in a spatially closed manifold $\mathcal{M}^D$:
	\begin{equation}
{M}= \mathcal{S}^1\times \mathcal{M}^D.
	\end{equation}
	As the gauge theory is compact, this implies flux quantization. According to the normalization used in the manuscript, this reads 
	\begin{equation}
	\int_{\mathcal{M}^D} B^{(\alpha)}_{a_1a_2\ldots a_{D-3}}\equiv \pi p^{(\alpha)}_{a_1a_2\ldots a_{D-3}}, ~~~p^{(\alpha)}_{a_1a_2\ldots a_{D-3}}\in \mathbb{Z},
	\label{flux}
	\end{equation}
	where $ B^{(\alpha)}_{a_1a_2\ldots a_{D-3}}$ is the magnetic field given in \eqref{ebgauge}, which we repeat here for convenience
	\begin{equation}
	B^{(\alpha)}_{a_1a_2\ldots a_{D-3}}=\epsilon_{a_1 a_2\ldots a_{D-1}}\mathcal{D}^{(\alpha)}_{a_{D-2}} A_{a_{D-1}}.
	\label{a}
	\end{equation}
	This implies that $p^{(\alpha)}_{a_1a_2\ldots a_{D-3}}$ are completely anti-symmetric. It is useful to invert relation \eqref{a}
	\begin{equation}
	\mathcal{D}^{(\alpha)}_{b_1}A_{b_2}-\mathcal{D}^{(\alpha)}_{b_2}A_{b_1}=\frac{1}{(D-3)!}\epsilon_{a_1a_2\ldots a_{D-3}b_1b_2}B^{(\alpha)}_{a_1a_2\ldots a_{D-3}}.
	\end{equation}
	
	Now, let us consider large gauge transformations that wind around the $\mathcal{S}^1$ (time direction)
	\begin{equation}
	\zeta^{(\alpha)}\equiv 2\pi n^{(\alpha)}\, \frac{t}{\tau},~~~n^{(\alpha)}\in\mathbb{Z}.
	\end{equation} 
	This implies
	\begin{equation}
	A_0^{(\alpha)}\rightarrow A_0^{(\alpha)}+2\pi n^{(\alpha)}\,\frac{1}{\tau}.
	\end{equation}
	The corresponding variation of the action is 
	\begin{eqnarray}
	\delta S&=&\int d^Dx dt \frac{1}{\pi} \sum_{\alpha} 2\pi n^{(\alpha)}\,\frac{1}{\tau} K_{ab}\mathcal{D}_a^{(\alpha)}A_b \nonumber\\
	&=&\int d^Dx dt  \sum_{\alpha}  n^{(\alpha)}\,\frac{1}{\tau} K_{ab}\left(\mathcal{D}_a^{(\alpha)}A_b-\mathcal{D}_b^{(\alpha)}A_a\right)\nonumber\\
	&=& \int \frac{dt}{\tau}\sum_{\alpha} n^{(\alpha)}\, K_{ab} \,\frac{1}{(D-3)!}\,\epsilon_{a_1a_2\ldots a_{D-3}a b}\, \int d^Dx\, B^{(\alpha)}_{a_1a_2\ldots a_{D-3}}.
	\end{eqnarray}
	By using the flux quantization (\ref{flux}), we obtain
	\begin{eqnarray}
	\delta S&=& \pi \sum_{\alpha} n^{(\alpha)}\, K_{ab} \frac{1}{(D-3)!}\;\epsilon_{a_1a_2\ldots a_{D-3}a b}\; p^{(\alpha)}_{a_1a_2\ldots a_{D-3}}.
	\label{1}
	\end{eqnarray}
	In order for the quantum theory to be invariant under large gauge transformations, we must have
	\begin{equation}
	\delta S = 2\pi i \mathbb{Z}.
	\label{2}
	\end{equation}
	Equation (\ref{1}) together with this condition imply the quantization of all elements of $K$. Let us choose, for example, that the only nonvanishing integers in (\ref{1}) are
	\begin{equation}
	n^{(1)}=1,~~~p^{(1)}_{3 4\ldots D-3}=1, ~~~\text{and the $(D-3)!$ permutations of}~ p^{(\alpha)}_{3 4\ldots D-3}.
	\label{3}
	\end{equation}
In this case, (\ref{1}) becomes
	\begin{eqnarray}
	\delta S &=&\pi K_{ab} \epsilon_{3 4 \ldots (D-3) a b} \nonumber\\
	&=& 2 \pi K_{12}. 
	\end{eqnarray}
Thus, $K_{12}$ must be an integer. By proceeding similarly we get the quantization of all elements of $K$. 
	
The flux quantization (\ref{flux}) also leads properly to the charge quantization. Indeed, by introducing the coupling to a density $-A_0^{(\alpha)}J_0^{(\alpha)}$, it follows the flux-attachment relation
	\begin{equation}
	J_0^{(\alpha)}=\frac{1}{\pi}  K_{ab}\mathcal{D}_a^{(\alpha)}A_b. 
	\end{equation}
By integrating over $\mathcal{M}^D$, we obtain
	\begin{eqnarray}
	Q^{(\alpha)}=\int_{\mathcal{M}^D} J_0^{(\alpha)}&=&\int_{\mathcal{M}^D} \frac{1}{\pi}  K_{ab}\mathcal{D}_a^{(\alpha)}A_b\nonumber\\
	&=&\frac{1}{2\pi} K_{ab} \int_{\mathcal{M}^D} \left(\mathcal{D}_a^{(\alpha)}A_b-\mathcal{D}_b^{(\alpha)}A_a \right)\nonumber\\
	&=&\frac{1}{2\pi} K_{ab} \int_{\mathcal{M}^D} \frac{1}{(D-3)!}\epsilon_{a_1a_2\ldots a_{D-3} a b}B^{(\alpha)}_{a_1a_2\ldots a_{D-3}}\nonumber\\
	&=& \frac{1}{2} K_{ab} \int_{\mathcal{M}^D} \frac{1}{(D-3)!}\epsilon_{a_1a_2\ldots a_{D-3} a b} \, p^{(\alpha)}_{a_1a_2\ldots a_{D-3}}\,.
	\end{eqnarray}
By choosing again the configuration in (\ref{3}), 
	\begin{eqnarray}
	Q^{(\alpha)}&=&\frac{1}{2} K_{ab} \epsilon_{3 4\ldots (D-3) a b}\nonumber\\
	&=& K_{12}\,,
	\end{eqnarray}
which is an integer.
	\section{$K$-matrix and microscopic theory}\label{AE}
	The matrix $K$ is determined in three steps: i) we determine its dimensionality; ii) we find constraints on the possible values of the elements; iii) we fix the elements in order to match the physical properties of the lattice model. The steps i) and ii) follow directly from the algebra of the operators in \eqref{eq2}. Step iii) is more subtle and follow from  the algebra of the ground state holonomy gauge-invariant operators of the effective field theory. 
	
\paragraph*{\textbf{Step (i): Dimensionality of $K$-matrix.}}
	
In the class of models we have constructed in the manuscript, the dimensionality of the ''spin" operator acting at each site is tied to the spatial dimensionality. Indeed, 
	\begin{eqnarray}
	D=3~~~&\Rightarrow&~~~\sigma_{i_1} \Leftrightarrow \gamma_{2\times 2}\nonumber\\
	D=5~~~&\Rightarrow&~~~\sigma_{i_1}\otimes \sigma_{i_2} \Leftrightarrow \gamma_{4\times 4}\nonumber\\
	D=2n+1 ~~~&\Rightarrow&~~~\sigma_{i_1}\otimes\sigma_{i_2}\otimes\cdots\otimes\sigma_{i_n} \Leftrightarrow \gamma_{2^n\times 2^n}.
	\end{eqnarray}
	In this way, the algebra of the spin operators in $D$ spatial dimensions can be written in terms of the Clifford algebra of Dirac matrices of dimensionality $2^n\times 2^n$. In this case, we have $2n$ Dirac matrices. Therefore, according to the representation of \eqref{eq2} of the paper, we need $2n$ distinct fields $A$ to reproduce properly the algebra of the Dirac matrices. This fixes the dimensionality of the matrix $K$ to be $2n \times 2n$.
	
\paragraph*{\textbf{Step (ii): Elements of the $K$-matrix.}}

	The possible values of the elements of the $K$-matrix are determined from the algebra of operators at each site. Starting in $D=3$, with the $K$-matrix given by
	\begin{equation}
	K=\left(
	\begin{array}{cc}
	0& k\\
	-k& 0\\
	\end{array}
	\right)
	~~~\text{and}~~~
	K^{-1}=\left(
	\begin{array}{cc}
	0& -\frac{1}{k}\\
	\frac{1}{k}& 0\\
	\end{array}
	\right),
	\end{equation}
	we have the two operators
	\begin{equation}
	\gamma^1=e^{i k A_2}~~~\text{and}~~~\gamma^2=e^{-i k A_1}.
	\end{equation} 
	They will anticommute if $k$ is odd. For the $D=2n+1$ dimensional case, the reasoning is similar. We consider a basis of fields so that the matrix $K$ is block-diagonal (eq.\eqref{eq67} of the manuscript)
		\vspace{-.08cm}
	\begin{equation}
	Q\,K\,Q^{T}=\text{Diag}\left\{
	\begin{pmatrix}
	0& k_1\\
	-k_1& 0
	\end{pmatrix}\,,
	\begin{pmatrix}
	0& k_2\\
	-k_2& 0
	\end{pmatrix}\,,\ldots\,,
	\begin{pmatrix}
	0& k_n\\
	-k_n& 0
	\end{pmatrix}
	\right\}\,.
	\label{eq67a}
	\end{equation}

	Then, the representation \eqref{eq2} of the paper will provide the properly commutation rules between the operators only if all $k$'s are odd.
	
\paragraph*{\textbf{Step (iii): Lattice model and the $K$-matrix entries.}}
	
To completely fix the $k_i$'s of the matrix in (\ref{eq67a}), we need to consider the algebra of gauge-invariant operators. This is worked in detail in Sec.\ref{GSD} of the manuscript  for the case $D=3$. The key equation is the algebra given in \eqref{eq55} of the paper. The value of $k_i$ determines the size of the representation of the ground state in a corresponding subdimensional manifold where charge is conserved. In other words, the ground state will possess a $\mathbb{Z}_{2k}$ symmetry that is not present in the lattice model unless $k=1$. The same reasoning goes in higher dimensions, where we have more pairs of operators satisfying the algebra \eqref{eq55} of the manuscript . Thus, the $K$-matrix corresponding to the lattice model is so that its block-diagonal form has $k_1=k_2=\ldots=k_n=1$. In this sense, the $K$-matrix is determined by the microscopic system since it carries information about the symmetries of the lattice model. 

The  equation \eqref{eq2a} of the manuscript suggest a class of equivalence for the $K$-matrices in our description, in a similar fashion as usual Chern-Simons theories that can lead to the same description with two different $K$-matrices. To make this point clear, consider a redefinition of the basis field in the effective action according to
\begin{equation}
A \rightarrow W A,
\end{equation}
where $W$ is a matrix with integer entries. This leads to a theory with new parameters
\begin{equation}
\tilde{K}= W^{\top} K W~~~\text{and}~~~\tilde{T}^{(I,\alpha)}= W^{-1}T^{(I,\alpha)}.
\label{trans}
\end{equation}
Thus, two effective theories with parameters $(K,\, T)$ and $(\tilde{K},\,\tilde{T})$ related through (\ref{trans}), with the matrix $W$ possessing integer entries and $\det W=1$, describes the same fracton system. Indeed, this implies
\begin{equation}
\text{Pf}(\tilde{K})=\text{Pf}(W^{\top} K W)=\text{Pf}(K)
\end{equation} 
and also leaves unchanged the quantization condition (for the principal configuration) given in \eqref{eq6} of the manuscript:
\begin{equation}
t^{(I)}_a \;(K^{\top})_{ab} \;\,t^{(J)}_b =\tilde{t}^{(I)}_a \;(\tilde{K}^{\top})_{ab} \;\,\tilde{t}^{(J)}_b= 2 n^{(IJ)}+(1-\delta_{IJ})\,,~~~  n^{(IJ)}\in\mathbb{Z}\,.
\end{equation}
which is still an even integer if $I=J$ and an odd integer if $I\neq J$.
	
	\section{Conservation Laws and Excitations}\label{AD}
	
	\subsection{Case $D=3$}
	
	We will examine here the possible types of defects arising from this model of fractons. Let us consider the case $D=3$, so that the continuity equation reads
	\begin{eqnarray}
	\partial_0\, J_0&=&\mathcal{D}_1\,J_1+\mathcal{D}_2\,J_2\nonumber\\
	&=&(\partial_x^2-\partial_z^2)\,J_1+(\partial_y^2-\partial_z^2)\,J_2\nonumber\\
	&=&(\partial_x+\partial_z)(\partial_x-\partial_z)\,J_1+(\partial_y+\partial_z)(\partial_y-\partial_z)\,J_2\nonumber\\
	&=& \partial^+_{13}\,\partial^-_{13}\, J_1+  \partial^+_{23}\,\partial^-_{23}\, J_2\,,
	\label{A33}
	\end{eqnarray} 
	where the coordinates are $x_{13}^{\pm}=x\pm z$ and $x_{23}^{\pm}=y\pm z$. For simplicity, we have absorbed a factor of $1/4$ in $J_0$.
	
	Let us try, for example, to construct a current corresponding to the creation of a single localized charge. For simplicity, we set $J_2=0$. Then, a naive solution of (\ref{A33}) is
	\begin{equation}
	J_0=\theta(t)\,\delta (y)\,\delta(x_{13}^++a_1)\,\delta(x_{13}^-+b_1)    ~~~\text{and}~~~J_1= \delta (t)\,\delta(y)\,\theta(x_{13}^++a_1)\,\theta(x_{13}^-+b_1)\,,
	\label{A34}
	\end{equation} 
	which corresponds to the creation of a fracton localized at $x=-(a_1+b_1)/2$, $y=0$ and $z=(b_1-a_1)/2$. Notice, however, that this configuration corresponds to a process where charge is not conserved $(Q=0\rightarrow 1)$. Indeed,
	\begin{equation}
	Q=\int dx dy dz \,J_0 = \theta(t)\int dy dx_{13}^+ dx_{13}^- \, \delta (y)\delta(x_{13}^++a_1)\delta(x_{13}^-+b_1)=\theta (t),~~~\frac{dQ}{dt}=\delta(t)\,.
	\label{A35}
	\end{equation}
	Consequently, it is not a full-fledged solution of the continuity equation.
	
	We could try to avoid the violation of charge above by inserting a charge of opposite sign in a distinct point, which corresponds
	\begin{eqnarray}
	J_0&=&\theta(t)\,\delta (y)\,\left[\delta(x_{13}^++a_1)\,\delta(x_{13}^-+b_1)-\delta(x_{13}^++c_1)\,\delta(x_{13}^-+d_1)\right]   \nonumber\\ 
	J_1&=& \delta (t)\,\delta(y)\,\left[\theta(x_{13}^++a_1)\,\theta(x_{13}^-+b_1)-\theta(x_{13}^++c_1)\,\theta(x_{13}^-+d_1)\right]\;.
	\label{A36}
	\end{eqnarray} 
	This is compatible with charge conservation in the whole system, $Q=\int dx \,dy\, dz\, J_0$, but we still have to inspect the conservation in the sub-manifolds. Let us consider, for example, the following charge
	\begin{eqnarray}
	Q^{(++)}&=&\int dx_{13}^{+}\,dx_{23}^{+}\, J_0\nonumber\\
	&=&\theta(t) \int dx_{13}^{+}\,dx_{23}^{+}\, \delta (y)\,\left[\delta(x_{13}^++a_1)\,\delta(x_{13}^-+b_1)-\delta(x_{13}^++c_1)\,\delta(x_{13}^-+d_1)\right]\;.
	\label{A37}
	\end{eqnarray}
	We need to be careful in computing the integrals, since the directions $x_{13}^{+}$ and $x_{23}^{+}$ are not orthogonal, whereas the directions $x_{13}^{+}$ and $x_{13}^{-}$ are orthogonal. This means that we can carry out the integration over $x_{13}^{+}$ keeping $x_{13}^{-}$ fixed. Thus, we proceed by integrating over $x_{13}^{+}$, letting $x_{13}^{-}$ untouched:
	\begin{eqnarray}
	Q^{(++)}&=&\theta(t)\, \int dx_{23}^{+}\, \delta (y)\,\left[\delta(x_{13}^-+b_1)-\delta(x_{13}^-+d_1)\right]\;.
	\label{A38}
	\end{eqnarray}
	The computation of the remaining integral is a little trick because the directions $x_{13}^{+}$ and $x_{23}^{+}$ are not orthogonal, but actually we do not need to compute it to extract useful information. Indeed, this expression shows that in order that the charge $Q^{(++)}$ to be conserved we need to require $b_1=d_1$. Similarly, by considering the charge 
	\begin{eqnarray}
	Q^{(-+)}&=&\theta(t) \,\int dx_{13}^{-}\,dx_{23}^{+} \,\delta (y)\,\left[\delta(x_{13}^++a_1)\,\delta(x_{13}^-+b_1)-\delta(x_{13}^++c_1)\,\delta(x_{13}^-+d_1)\right]\nonumber\\
	&=&\theta(t)  \int dx_{23}^{+}\, \delta (y)\,\left[\delta(x_{13}^++a_1)-\delta(x_{13}^++c_1)\right]\;,
	\label{A39}
	\end{eqnarray}
	we see that $a_1=c_1$ in order that this charge to be conserved. The charges $Q^{(+-)}$ and $Q^{(--)}$ do not provide additional conditions. Taking into account that $a_1=c_1$ and $b_1=d_1$ in (\ref{A36}), we see that the density of charges $J_0$ trivially vanishes. In conclusion, the process of creation of a dipole is not compatible with the several conservation laws and, consequently, it is not allowed.

\begin{figure}[!h]
		\centering
		\includegraphics[scale=0.7]{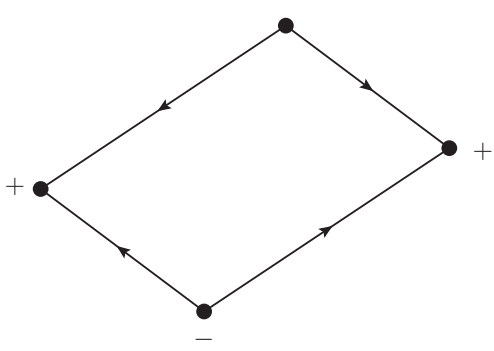}
		\caption{A generic four-charge configuration in the plane $x-z$, as given in (\ref{A46}). It is clear from this figure that the total dipole vanishes.}
		\label{genfrac}
	\end{figure}

	Let us try to find a different type of configuration, which is compatible with the whole set of conservation laws. Consider the density,
	\begin{eqnarray}
	J_0&=&\theta(t)\,\delta(y)\,\left[\delta(x_{13}^++a_1)\,\delta(x_{13}^{-}+b_1)-\delta\,(x_{13}^++c_1)\,\delta(x_{13}^-+d_1)\right.\nonumber\\&-&\left.\delta(x_{13}^++e_1)\,\delta(x_{13}^-+f_1)+\delta(x_{13}^++g_1)\,\delta(x_{13}^-+h_1) \right]\;,
	\label{A40}
	\end{eqnarray}
	and the corresponding flux
	\begin{eqnarray}
	J_1&=&\delta(t)\,\delta(y)\,\left[\theta(x_{13}^++a_1)\,\theta(x_{13}^{-}+b_1)-\theta(x_{13}^++c_1)\,\theta(x_{13}^-+d_1)\right.\nonumber\\&-&\left.\theta(x_{13}^++e_1)\,\theta(x_{13}^-+f_1)+\theta(x_{13}^++g_1)\,\theta(x_{13}^-+h_1) \right]\;,
	\label{A40a}
	\end{eqnarray}
	which are compatible with the continuity equation (\ref{A33}). It follows immediately that the charge is conserved in the whole three-dimensional manifold. Next, let us examine the conservation laws in the sub-manifolds. We start with the following charges,
	\begin{eqnarray}
	Q^{(+\pm)}&=&\int dx_{13}^{+} \,dx_{23}^{\pm}\, J_0\nonumber\\
	&=& \theta(t)\int dx_{23}^{\pm}\left[\delta(x_{13}^{-}+b_1)-\delta(x_{13}^-+d_1)-\delta(x_{13}^-+f_1)+\delta(x_{13}^-+h_1) \right]\;.
	\label{A41}
	\end{eqnarray}
	We have two possibilities ensuring charge conservation:
	\begin{eqnarray}
	&&i)~ b_1=d_1~~~\text{and}~~~ f_1=h_1\nonumber\\
	&&ii)~b_1=f_1~~~\text{and}~~~ d_1=h_1.
	\label{A42}
	\end{eqnarray}
	Similarly, the remaining charges are 
	\begin{eqnarray}
	Q^{(-\pm)}&=&\int dx_{13}^{-} \,dx_{23}^{\pm}\, J_0\nonumber\\
	&=& \theta(t)\int dx_{23}^{\pm}\left[\delta(x_{13}^++a_1)-\delta(x_{13}^++c_1)-\delta(x_{13}^++e_1)+\delta(x_{13}^++g_1) \right]\,,
	\label{A43}
	\end{eqnarray}
	which leads also to two possibilities
	\begin{eqnarray}
	&&i)~ a_1=c_1~~~\text{and}~~~ e_1=g_1\nonumber\\
	&&ii)~a_1=e_1~~~\text{and}~~~ c_1=g_1\;.
	\label{A44}
	\end{eqnarray}
	From these possibilities, it is clear that if we select choice $i)$ of (\ref{A42}) and $i)$ of (\ref{A44}), or $ii)$ of (\ref{A42}) and $ii)$ of (\ref{A44}), the density in (\ref{A40}) will trivially vanish. However, we obtain a non-vanishing density if we choose crosswise $i)/ ii)$ of (\ref{A42}) and $ii)/i)$ of (\ref{A44}). Let us choose, say, $i)$ from (\ref{A42}) and $ii)$ from (\ref{A44}). In this case, the density becomes
	\begin{eqnarray}
	J_0&=&\theta(t)\,\delta(y)\,\left[\delta(x_{13}^++a_1)\,\delta(x_{13}^{-}+b_1)-\delta(x_{13}^++c_1)\,\delta(x_{13}^-+b_1)\right.\nonumber\\&-&\left.\delta(x_{13}^++a_1)\,\delta(x_{13}^-+f_1)+\delta(x_{13}^++c_1)\,\delta(x_{13}^-+f_1) \right]\;,
	\label{A45}
	\end{eqnarray}
	which corresponds to the creation of four charges at the following positions:
	\begin{eqnarray}
	\text{charge}~&q_1=+&~\Rightarrow~(x,z)=\left(-\frac{a_1+b_1}{2}, \frac{b_1-a_1}{2}\right)\nonumber\\
	\text{charge}~&q_2=-&~\Rightarrow~(x,z)=\left(-\frac{c_1+b_1}{2}, \frac{b_1-c_1}{2}\right)\nonumber\\
	\text{charge}~&q_3=-&~\Rightarrow~(x,z)=\left(-\frac{a_1+f_1}{2}, \frac{f_1-a_1}{2}\right)\nonumber\\
	\text{charge}~&q_4=+&~\Rightarrow~(x,z)=\left(-\frac{c_1+f_1}{2}, \frac{f_1-c_1}{2}\right)\;.
	\label{A46}
	\end{eqnarray}
	Let $d(q_i,q_j)$ be the distance between two charges. The above expressions ensure that $d(q_1,q_2)=d(q_3,q_4)$ and $d(q_1,q_3)=d(q_2,q_4)$, which physically means that the sum of all dipole moments of the configuration vanishes (see figure \ref{genfrac}). This guarantees conservation of dipole moment, which is a consequence of the conservation of charges in sub-manifolds (planes). In fact, charge conservation along a plane implies that the dipole moment perpendicular to the plane is conserved.  
	
	We see that the location of the four charges in (\ref{A46}) are specified by the set of arbitrary points $a_1,b_1,c_1,f_1$. By varying the values of these points we change both the size of the dipoles and their positions, in a way that preserves the structure depicted in figure \ref{genfrac}, i.e., the charges are always localized at the corners of a parallelogram. Physically, this means that the dipoles can move freely in the system, but cannot be created or annihilated (remembering our previous discussion, the creation of a single dipole is not compatible with all the conservation laws).
	
\begin{figure}[!h]
		\centering
		\includegraphics[scale=0.7]{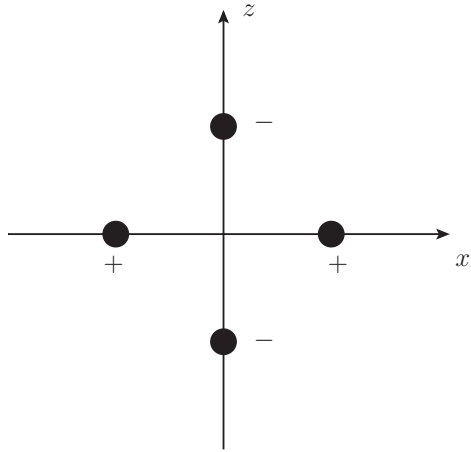}
		\caption{Charge configuration corresponding to the process described by the density in (\ref{A36}).}
		\label{fracton}
	\end{figure}

	Before closing, it is instructive to consider a simple symmetric choice, $a_1=b_1=a$ and $c_1=f_1=-a$.  In this case, the density reduces to 
	\begin{eqnarray}
	J_0&=&\theta(t)\,\delta(y)\left[\delta(x_{13}^++a)\,\delta(x_{13}^-+a)-\delta(x_{13}^+-a)\,\delta(x_{13}^-+a)\right.\nonumber\\&-&\left.\delta(x_{13}^++a)\,\delta(x_{13}^--a)+\delta(x_{13}^+-a)\,\delta(x_{13}^--a) \right]\,,
	\label{A47}
	\end{eqnarray}
	while the flux can be written as 
	\begin{eqnarray}
	J_1&=&\delta(t)\,\delta(y)\left[\theta(x_{13}^++a)\,\theta(x_{13}^{-}+a)-\theta(x_{13}^+-a)\,\theta(x_{13}^-+a)\right.\nonumber\\&-&\left.\theta(x_{13}^++a)\,\theta(x_{13}^--a)+\theta(x_{13}^+-a)\,\theta(x_{13}^--a) \right]\nonumber\\
	&=& \delta (t)\,\delta(y)\,\theta(a+x_{13}^+)\,\theta(a-x_{13}^+)\,\theta(a+x_{13}^-)\,\theta(a-x_{13}^-)\,,
	\label{A48}
	\end{eqnarray} 
	where to write in terms of a single term we have used the property $\theta(x)+\theta(-x)=1$. The density $J_0$ describes the creation of a set of four charges located at the points
	\begin{eqnarray}
	x=\pm a,~y=0,~z=0~~~&&\Rightarrow ~~~\text{positive charges}\nonumber\\
	x=0,~y=0,~z=\pm a~~~&&\Rightarrow ~~~\text{negative charges}\;.
	\label{A49}
	\end{eqnarray}
	This configuration is depicted in figure \ref{fracton}. 

	\subsection{Case $D=5$}	
	
	In this case we have two conservation laws given by (\ref{eq36}), 
	\begin{equation}
	\partial_0 \,J_0^{(1)}=D_1\, J_1^{(1)}+D_2\,J_2^{(1)}+D_3\,J_3^{(1)}+D_4\,J_4^{(1)}\;,
	\label{A50}
	\end{equation}
	and 
	\begin{equation}
	\partial_0 \,J_0^{(2)}=D_1\, J_1^{(2)}+D_2\,J_2^{(2)}+D_3\,J_3^{(2)}+D_4\,J_4^{(2)}\;,
	\label{A51}
	\end{equation}
	where $J_I^{(\alpha)}=T_a^{(I,\alpha)}\,J_a$. For $\alpha=1$ and the canonical form of $t$'s,
	\begin{equation}
	J_I^{(1)}=t_a^{(I)}\,J_a=J_I\;,
	\label{A52}
	\end{equation}	
	whereas for $\alpha=2$ it is convenient to write all components explicitly,
	\begin{eqnarray}
	J_1^{(2)}&=&T_1^{(1,2)}\, J_1 +T_2^{(1,2)}\, J_2+T_3^{(1,2)}\, J_3+T_4^{(1,2)}\, J_4\nonumber\\
	J_2^{(2)}&=&T_1^{(2,2)}\, J_1 +T_2^{(2,2)}\, J_2+T_3^{(2,2)}\, J_3+T_4^{(2,2)}\, J_4\nonumber\\
	J_3^{(2)}&=&T_1^{(3,2)}\, J_1 +T_2^{(3,2)}\, J_2+T_3^{(3,2)}\, J_3+T_4^{(3,2)}\, J_4\nonumber\\
	J_4^{(2)}&=&T_1^{(4,2)}\, J_1 +T_2^{(4,2)}\, J_2+T_3^{(4,2)}\, J_3+T_4^{(4,2)}\, J_4\;.
	\label{A53}
	\end{eqnarray}
	
	Our goal here is the following. We will create an elementary excitation in the lattice model and then we want to understand how this is reproduced from the conservation laws above. Let us consider the case $D=5$ given in (\ref{d5}). The application of the local operator $\gamma^1\,\gamma^2$ in a particular site of the even sub-lattice creates four defects of type $\alpha=1$ in the plane $x_1-x_2$ and four defects of type $\alpha=2$ in the plane $x_3-x_4$. Now let us see how this arises from the point of view of the conservation laws. 
	
	These excitations can be reproduced with $J_3=J_4=0$ and $J_2=-J_1$, so that (\ref{A50}) becomes
	\begin{eqnarray}
	\partial_0 J_0^{(1)}&=&(D_1-D_2)J_1\nonumber\\
	&=&(\partial_1^2-\partial_2^2) J_1\nonumber\\
	&=&\partial_{12}^+ \,\partial_{12}^-\, J_1\;.
	\label{A54}
	\end{eqnarray}
	To construct the currents for $\alpha=2$ we can read the vectors $T$ from $(\ref{B30})$. We remember, however, that their components are defined only mod 2, so that it is convenient to choose	
	\begin{eqnarray}
	J_1^{(2)}&=&J_2^{(2)}~~=0\nonumber\\
	J_3^{(2)}&=&-J_4^{(2)}=J_1\;.
	\label{A55}
	\end{eqnarray}
	With this, the conservation law (\ref{A51}) becomes
	\begin{equation}
	\partial_0 \,J_0^{(2)}=\partial_{34}^+ \,\partial_{34}^- \,J_1\;.
	\label{A56}
	\end{equation}
	We can construct a current $J_1$ that creates excitations simultaneously in planes $x_1-x_2$ and $x_3-x_4$ by using the four-charge configurations of the case $D=3$ (\ref{A40a}) with the positions of the charges subject to (\ref{A46}), since these configurations also live in planes. In this way, we write the generalization for the five-dimensional case as
	\begin{equation}
	J_1=\delta(t)\,\delta(x_3)\,\delta(x_4)\,\delta(x_5)\,\Theta(x_{12}^+,x_{12}^-)+\delta(t)\,\delta(x_1)\,\delta(x_2)\,\delta(x_5)\,\Theta(x_{34}^+,x_{34}^-)\,,
	\label{A57}
	\end{equation}
	where $\Theta(x_{12}^+,x_{12}^-)$ is defined as the $\theta$-dependent part of (\ref{A40a}):
	\begin{eqnarray}
	\Theta(x_{12}^+,x_{12}^-)&\equiv& \theta(x_{12}^++a_1)\,\theta(x_{12}^{-}+b_1)-\theta(x_{12}^++c_1)\,\theta(x_{12}^-+b_1)\nonumber\\&-&\theta(x_{12}^++a_1)\,\theta(x_{12}^-+f_1)+\theta(x_{12}^++c_1)\,\theta(x_{12}^-+f_1)\,.
	\label{A58} 
	\end{eqnarray} 
	Plugging $J_1$ in (\ref{A54}) and (\ref{A56}) gives the densities
	\begin{eqnarray}
	J_0^{(1)}&=&\theta(t)\, \delta(x_3)\,\delta(x_4)\,\delta(x_5)\,\Delta(x_{12}^+,x_{12}^-)\nonumber\\&+& \theta(t)\,\delta(x_5)\,\Theta(x_{34}^+,x_{34}^-)\, \partial_{12}^+\,\partial_{12}^-\delta(x_1)\,\delta(x_2)\;,
	\label{A59}
	\end{eqnarray}
	and 
	\begin{eqnarray}
	J_0^{(2)}&=&\theta(t)\,\delta(x_1)\,\delta(x_2)\,\delta(x_5)\,\Delta(x_{34}^+,x_{34}^-)\nonumber\\&+&\theta(t)\,\delta(x_5)\,\Theta(x_{12}^+,x_{12}^-)\,\partial_{34}^+\,\partial_{34}^-\, \delta(x_3)\,\delta(x_4)\;,
	\label{A60}
	\end{eqnarray}
	where 
	\begin{eqnarray}
	\Delta(x_{12}^+,x_{12}^-)&\equiv& \partial_{12}^+\,\partial_{12}^-\,\Theta(x_{12}^+,x_{12}^-) \nonumber\\&=& \delta(x_{12}^++a_1)\,\delta(x_{12}^{-}+b_1)-\delta(x_{12}^++c_1)\,\delta(x_{12}^-+b_1)\nonumber\\&-&\delta(x_{12}^++a_1)\,\delta(x_{12}^-+f_1)+\delta(x_{12}^++c_1)\,\delta(x_{12}^-+f_1)\,.
	\label{A61} 
	\end{eqnarray} 
	There are some important points to notice in the densities $J_0^{(1)}$ and $J_0^{(2)}$. The terms in the first lines of both (\ref{A59}) and (\ref{A60}) correspond indeed to four-charge configurations with vanishing total dipole, like in the case $D=3$. But now, we have additional terms in the second lines. However, such terms do not affect the physical charge and can be absorbed in a redefinition of the currents. Indeed, we can define
	\begin{equation}
	\tilde{J}_0^{(1)}\equiv J_0^{(1)}-\Omega_0^{(1)}~~~\text{and}~~~ \tilde{J}_1^{(1)}\equiv J_1-\Omega_1^{(1)}\;. 
	\label{A61a}
	\end{equation}
	with similar definitions for the currents of $\alpha=2$, i.e., $\tilde{J}_0^{(2)}\equiv  J_0^{(2)}-\Omega_0^{(2)} $ and $J_1^{(2)}\equiv J_1-\Omega_1^{(2)}$.  
	If $\Omega_0^{(1)}$ and $\Omega_1^{(1)}$ satisfy
	\begin{equation}
	\partial_0\, \Omega_0^{(1)}= \partial_{12}^+\,\partial_{12}^-\, \Omega_1^{(1)}\;,
	\label{A61b}
	\end{equation}
	and
	\begin{equation}
	\int  dx_{15}^{\sigma_1}\, dx_{25}^{\sigma_2} \,dx_{35}^{\sigma_3}\, dx_{45}^{\sigma_4}\; \Omega_0^{(1)}=0,
	\label{A61c}
	\end{equation}
	then the two currents $(J_0^{(1)},\, J_1)$ and $(\tilde{J}_0^{(1)}, \,\tilde{J}_1^{(1)})$ describe the same physical situation, since the redefined currents also satisfy
	\begin{equation}
	\partial_0 \,\tilde{J}_0^{(1)}=\partial_{12}^+\,\partial_{12}^-\, \tilde{J}_1^{(1)}\,,
	\label{A61d}
	\end{equation}
	and
	\begin{equation}
	\int  dx_{15}^{\sigma_1}\, dx_{25}^{\sigma_2}\, dx_{35}^{\sigma_3}\, dx_{45}^{\sigma_4}\, \tilde{J}_0^{(1)}=\int  dx_{15}^{\sigma_1}\, dx_{25}^{\sigma_2}\, dx_{35}^{\sigma_3}\, dx_{45}^{\sigma_4}\, {J}_0^{(1)}\;.
	\label{A61e}
	\end{equation}

	From equations (\ref{A57}) and (\ref{A59}) we see that if we  set,
	\begin{equation}
	\Omega_0^{(1)}=\theta(t)\,\delta(x_5)\,\Theta(x_{34}^+,x_{34}^-) \partial_{12}^+\,\partial_{12}^-\,\delta(x_1)\,\delta(x_2)\;,
	\label{A61d}
	\end{equation}
	and 
	\begin{equation}
	\Omega_1^{(1)}=\delta(t)\,\delta(x_1)\,\delta(x_2)\,\delta(x_5)\,\Theta(x_{34}^+,x_{34}^-)\,,
	\label{A61e}
	\end{equation}
	then the condition (\ref{A61b}) is immediately satisfied. 
	
	Next, let us consider (\ref{A61c}),
	\begin{equation}
	\int dx_{15}^{\sigma_1}\, dx_{25}^{\sigma_2}\, dx_{35}^{\sigma_3}\, dx_{45}^{\sigma_4}\; \theta(t)\,\delta(x_5)\,\Theta(x_{34}^+,x_{34}^-)\, (\partial_1^2-\,\partial_2^2)\,\delta(x_1)\,\delta(x_2)\;.
	\label{A62}
	\end{equation}
	This term vanishes identically. To see this, we notice that as $x_{15}^{\sigma_1}=x_1+\sigma_1\, x_5$ and $x_{25}^{\sigma_2}=x_2+\sigma_2\, x_5$, with $\sigma_1,\sigma_2=\pm$, under the change of variables
	\begin{equation}
	x_1 \rightarrow \sigma_1\,\sigma_2\, x_2 ~~~\text{and}~~~x_2 \rightarrow \sigma_1\,\sigma_2\, x_1\;,
	\label{A63}
	\end{equation}
	the integration measure transforms as 
	\begin{equation}
	dx_{15}^{\sigma_1}\rightarrow \sigma_1\,\sigma_2\, dx_{25}^{\sigma_2}  ~~~\text{and}~~~dx_{25}^{\sigma_2}\rightarrow \sigma_1\,\sigma_2\, dx_{15}^{\sigma_1}\;,
	\label{A64}
	\end{equation}
	so that $dx_{15}^{\sigma_1}\,dx_{25}^{\sigma_2}$ is invariant (even). On the other hand, the integrand $(\partial_1^2-\partial_2^2)\,\delta(x_1)\,\delta(x_2)$ is odd and hence the integral vanishes. Therefore, we can construct a redefined density simply as
	\begin{equation}
	\tilde{J}_0^{(1)}=\theta(t)\, \delta(x_3)\,\delta(x_4)\,\delta(x_5)\,\Delta(x_{12}^+,x_{12}^-)\;,
	\label{A65}
	\end{equation}
	which corresponds to the creation of a four-charge configuration in the plane $x_1-x_2$. We can proceed in the same way for the density in (\ref{A60}), and define 
	\begin{equation}
	\tilde{J}_0^{(2)}=\theta(t)\,\delta(x_1)\,\delta(x_2)\,\delta(x_5)\,\Delta(x_{34}^+,x_{34}^-)\;.
	\label{A66}
	\end{equation}
	It remains to show that these densities satisfy the requirement of charge conservation. This is not immediate because the four-charge configurations $\Delta(x_{12}^+,x_{12}^-)$ and $\Delta(x_{34}^+,x_{34}^-)$ involve directions which are not appearing in the integration measure (\ref{charge2}). For example, consider the charge
	\begin{eqnarray}
	Q^{(1)}_{(\sigma_1,\sigma_2,\sigma_3,\sigma_4)}&=&\int dx_{15}^{\sigma_1}\, dx_{25}^{\sigma_2}\, dx_{35}^{\sigma_3}\, dx_{45}^{\sigma_4}\, \tilde{J}_0^{1}
	\nonumber\\&=& \theta(t)\,\int dx_{15}^{\sigma_1}\, dx_{25}^{\sigma_2}\, dx_{35}^{\sigma_3}\, dx_{45}^{\sigma_4}\, \delta(x_3)\,\delta(x_4)\,\delta(x_5)\,\Delta(x_{12}^+,x_{12}^-)\;.
	\label{A67}
	\end{eqnarray}
	We have to change the integration from $x_{15}^{\pm}$ to $x_{12}^{\pm}$, since we know that $\int dx_{12}^{\sigma_1}\, \Delta(x_{12}^+,x_{12}^-)=0$. This can be done in the following way:
	\begin{eqnarray}
	x_{15}^{\sigma_1}&=&x_1+\sigma_1\, x_5\,,\nonumber\\
	&=&x_1+\tilde{\sigma}_1 \,x_2 + \sigma_1\, x_5-\tilde{\sigma}_1\, x_2\,,\nonumber\\
	&=& x_{12}^{\tilde{\sigma}_1}-\tilde{\sigma}_1 \,x_{25}^{-\tilde{\sigma}_1\sigma_1}\;.
	\label{A68}
	\end{eqnarray}
	As the change from $x_{15}^{\sigma_1}$ to $x_{12}^{\tilde{\sigma}_1}$ involves $x_{25}^{\pm}$, we have to ensure that the coordinate appearing in this expression is the opposite to the coordinate in the integration measure $dx_{25}^{\sigma_2}$, since the directions $x_{25}^+$ and $x_{25}^-$ are orthogonal. To this, we just need to set $\tilde{\sigma}_1=\sigma_1\sigma_2$, 
	\begin{equation}
	x_{15}^{\sigma_1}=x_{12}^{\sigma_1\sigma_2}-\sigma_1\,\sigma_2\, x_{25}^{-\sigma_2}\;.
	\label{A69}
	\end{equation}
	Therefore, as $x_{25}^{-\sigma_2}$ is fixed in the integration along the direction $x_{25}^{\sigma_2}$, we can directly write $dx_{15}^{\sigma_1}=dx_{12}^{\sigma_1\sigma_2}$, so that
	\begin{eqnarray}
	Q^{(1)}_{(\sigma_1,\sigma_2,\sigma_3,\sigma_4)}=\theta(t)\int dx_{12}^{\sigma_1}\, dx_{25}^{\sigma_2}\, dx_{35}^{\sigma_3}\, dx_{45}^{\sigma_4}\, \delta(x_3)\,\delta(x_4)\,\delta(x_5)\,\Delta(x_{12}^+,x_{12}^-)=0\,,
	\label{A70}
	\end{eqnarray}
	where we have renamed $\sigma_1 \,\sigma_2\rightarrow \sigma_1$. The same reasoning can be done with the charges associated with the density $\tilde{J}_0^{(2)}$. 
	
\end{appendix}

\bibliographystyle{unsrt}
\bibliography{FractonCliffFinal.bib}

		
\end{document}